\documentclass[useAMS,usenatbib]{mn2e}
\usepackage{psfig}
\usepackage{graphicx}
\pdfoutput=1

\def\kmskpc{{\rm ~km~s^{-1}~kpc^{-1}}}
\def\msol{{\rm ~M_\odot}}
\def\kms{{\rm km~s^{-1}}}

\def\Ro{R_{{o}}}
\def\mas{\textrm{mas yr}^{-1}}

\title[The Sagittarius Stream and Halo Triaxiality]
{The Sagittarius Stream and Halo Triaxiality}
\author[N. Deg \& L.Widrow]
  {{Nathan Deg$^{1}$\thanks{E-mail:ndeg@astro.queensu.ca},
  Lawrence Widrow$^{1}$}\\
  $^1$Department of Physics, engineering Physics, and Astronomy,Queen's University, 
\\Kingston, ON, K7L 3N6, Canada}

\begin{document}

\maketitle

\label{firstpage}

\begin{abstract}
We present a mass model for the Milky Way, which is fit to
observations of the Sagittarius stream together with constraints derived
from a wide range of photometric and kinematic data.  The model
comprises a S\'{e}rsic bulge, an exponential disk, and an Einasto halo.
Our Bayesian analysis is accomplished using an affine-invariant Markov
chain Monte Carlo algorithm.  We find that the best-fit dark matter
halo is triaxial with axis ratios of $3.3\pm 0.7$ and $2.7\pm 0.4$
and with the short axis approximately aligned with the Sun-Galactic centre
line.  Our results are consistent with those presented in Law and
Majewski (2010).  Such a strongly aspherical halo is disfavoured by
the standard cold dark matter scenario for structure formation.  
\end{abstract}
 

\section{Introduction}

It is a prediction of the hierarchical clustering scenario, borne out
by N-body simulations, that dark matter halos are triaxial (see, for
example, \citet{Frenk1988}, \citet*{Franx1991}, \citet{Warren1992},
\citet{Jing2002}, and \citet{Allgood2006}).  To be sure, baryons will
alter the shapes of halos, presumably making them more spherical
\citep{Dubinski1994, Kazantzidis2004, DeBattista2008}.  Nevertheless,
halo triaxiality is expected to be a feature of our Universe if
structure formation proceeds according to the standard model.
Unfortunately, the shapes of dark matter halos are extremely difficult
to determine by direct observations.

Tidal debris from the Milky Way's satellite galaxies provide a
potentially powerful probe of the Galactic gravitational potential
with the tidal stream from the Sagittarius dwarf spheroidal galaxy
(Sgr dSph) the most prominent example.  In the years following the
discovery of the Sgr dSph \citep{Ibata1997} numerous surveys have
produced an all-sky panorama of the Sgr system (see
\citet{Majewski2003} and references therein).  The observed positions
and velocities of stars in the stream are compared with theoretical
predictions for a range of triaxial halo models.  The underlying
assumption is that the kinematics of the stream is determined
primarily by the Galactic potential, and only to a lesser extent, by
the structure of the progenitor.  Examples of studies that attempted
to use the Sgr system in this way include \citet{Ibata2001},
\citet{Helmi2004B}, \citet{Martinez2004}, \citet*{Johnston2005}, and
\citet{Fellhauer2006}.

\citet{Majewski2003} presented an all-sky map of the Sgr system in
M-giant stars in which candidate Sgr stars were selected from the
2MASS catalog using their color, magnitude, and angular position.  The
result was a kinematic snapshot of both leading and trailing streams.
\citet{Belokurov2006} provided a more refined map of the stream using
the SDSS.  Recently \citet*{Law2009} and \citet{Law2010} (hereafter
LMJ09 and LM10, respectively) used both of these data sets to argue
that the Milky Way's halo is triaxial.  LMJ09 compared single particle
orbits to the projected positions of the SDSS fields and radial
velocities of the M giants.  Their best-fit model has isopotential
axis ratios of $q_{1,\Phi} = 1.5$ and $q_{z,\Phi}=1.25$ with the
halo's intermediate axis pointing toward the North Galactic pole and
its minor axis roughly aligned with the line that connects the Sun and
the Galactic centre (GC).  LM10 carried out a similar analysis, but
modeled Sagittarius as an N-body system rather than a single particle.
Their favored halo model is nearly oblate with $q_{1,\Phi}\simeq
q_{z,\Phi}\simeq 1.4$ where again the minor axis of the halo lies in
the disk plane nearly along the Sun-GC line.

  The LMJ09 and LM10 results present a challenge for the standard cold
  dark matter model of structure formation.  In general, a halo's
  gravitational potential will be more spherical than its mass
  distribution; the isodensity and isopotential axis ratios of an
  axisymmetric system ($q_\rho$ and $q_\Phi$ respectively) roughly
  satisfy the relation $q_{\rho}\simeq 3\left (q_\Phi-1\right) + 1$,
  at least in the outer parts of the halo \citep{BT2008}.  Thus, the
  mass distribution that corresponds to the LM10 potential is
  approximately oblate with axis ratios $\simeq 2.2$.  Since the
  formation of the Galactic disk is expected to make the halo more
  symmetric in the disk plane the axis ratios of the proto-galaxy are
  likely even larger. \citet{Allgood2006} presented a comprehensive
  analysis of dark halo shapes based on high-resolution
  dissipationless simulations.  They found that the mean
  major-to-minor isodensity axis ratio for Galaxy-sized halos was
  $q_\rho\simeq 1.7$ with a dispersion of $\sigma_q\simeq 0.3$.
  Moreover their halos, and those of earlier studies, tended more
  toward prolate-triaxial than oblate-triaxial (that is triaxial with
  one axis significantly smaller than the other two).  The effect
  baryons have on a halo's shape was studied by \citet{Dubinski1994,
    Kazantzidis2004, DeBattista2008} and others.  In general baryons
  tend to make the halos more spherical, particularly in the innermost
  regions.  The axis ratios of the proto-galaxy should therefore be
  even larger than those observed today.  The implication is that the
  Galactic halo favoured by the LMJ09 and LM10 analyses is an outlier
  within the context of the standard cosmological paradigm.

  A secondary issue concerns the shapes of the isodensity contours
  themselves.  LMJ09 and LM10 assume triaxial isopotential contours,
  which imply peanut-shaped isodensity contours (see, for example,
  \citet{BT2008}).  By contrast, triaxiality in simulations is almost
  always measured in terms of the density and peanut-shaped contours
  are rarely seen.  
 
  These considerations motivated us to reexamine constraints on
  the dark halo from the Sgr stream.  We carry out the
  modeling exercise within the Bayesian statistical framework and are
  therefore able to calculate the probability distribution function
  (PDF) for the structural parameters of the halo and quantify the
  model uncertainties.  Previous analyses focused on the shape
  parameters of the halo (the axis ratios and the Euler angles that
  define the halo's orientation).  The general practice has been to
  fix the structural parameters of the bulge, disk, and spherically-averaged 
  halo and allow the halo shape parameters to vary while
  trying to fit kinematic data for stars along the Sgr stream.
  Typically, only the most rudimentary constraints on the Galaxy's
  structure are included.  For example, in the LMJ09 and LM10 analyses
  (see, also \citet{Johnston2005}) the model is designed so that the
  circular speed at the position of the Sun is $v_c \simeq 220\,{\rm
    km}\,{\rm s}^{-1}$.  As discussed below, the disk and bulge in
  this model appear to be too large (by factors of 2-3) to be
  consistent with a number of observational constraints.

  In this paper, we consider a general disk-bulge-halo model for the
  Milky Way and allow the key structural parameters for all three
  components to vary simultaneously.  The likelihood function includes
  not only the stream constraints, but observational constraints from
  the line-of-sight velocity dispersion (LOSVD) and surface brightness
  profiles for the bulge, the vertical force and surface density in
  the solar neighborhood, the Oort constants, the circular speed
  curve, and the mass at large Galactocentric radii.

  At first glance, our parameter space has expanded to an unwieldy
  level.  (Our model has twenty-eight free parameters!).  However, by
  using a Markov chain Monte Carlo (MCMC) algorithm we are able to
  efficiently estimate the PDF for the full multi-dimensional
  parameter space.  In this paper, we employ the affine-invariant
  'Stretch-Move' ensemble sampler of \citet{Goodman2010} (see also
  \citet{Foreman-Mackey2012}).  Since our
  MCMC analysis requires a large number of ``likelihood calls'' we are
  precluded from modeling Sgr as a full N-body system.  We therefore
  follow LMJ09 and model Sgr and the associated tidal stream as a
  single particle that orbits in the fixed gravitational potential of
  the Milky Way.  The stream likelihood function is calculated by
  comparing the phase space coordinates from the M giant and SDSS
  observations (namely, the angular positions and line-of-sight
  velocities) with points along model orbits.

  In Section 2 we use Bayesian inference and our MCMC algorithm to
  reanalyze the M giant and SDSS data within the context of the
  Galactic model from LMJ09 and LM10.  We also consider an alternative
  model in which the isodensity contours (rather than isopotential
  contours) of the halo are triaxial.  In Section 3 we introduce a
  more general model for the density-potential pair of the Galaxy as
  well as the observational constraints that enter our analysis.  We
  present our main results in Section 4.  Our analysis points to a
  similarly triaxial model as the one found in LM10 though the
  disk-halo decomposition of the mass distribution is quite different.
  We conclude in Section 5 with a summary of our key results.

\section{Halo triaxiality from the Sagittarius stream via Bayesian inference}

We begin this section with brief summaries of the Galactic model used in
LMJ09 and LM10, the observational constraints from the Sgr stream, and
the affine-invariant ensemble sampler deployed in our analysis.  We
then present results from two separate MCMC runs.  The first closely
follows LMJ09 in that the isopotential contours are assumed to be
triaxial ellipsoids and certain model parameters, namely the
present-day distance to the Sgr dSph and the thickness and intrinsic
dispersion of the stream are kept fixed.  In the second run, the
isodensity contours are assumed to be triaxial and the aforementioned
parameters are allowed to vary.

\subsection{Galactic model}

The Galactic model used by LMJ09 and LM10 comprises a Hernquist bulge
\citep{Hernquist1990}, a Miyamoto-Nagai disk \citep{Miyamoto1975}, and
a logarithmic halo (see, for example, \citet{BT2008}).  The
contributions to the gravitational potential for each of these
components are given by
\begin{equation}\label{eq:LJMBulgePot} \Phi_{b}(r) ~=~-\frac{GM_{b}}{r+c} ~,\end{equation} 
\begin{equation}\label{eq:LJMDiskPot}
  \Phi_{d}(R,z)~=~-\frac{GM_{d}}{\sqrt{R^{2}+(a+\sqrt{z^{2}+b^{2}})^{2}}} ~,\end{equation} 
\begin{equation}\label{eq:LJMHaloPot}
  \Phi_{h}(r_{t})~=~\sigma_{h}^{2}\ln(r_{t}^{2}+d^{2})~.
\end{equation}
where $\left (x,\,y,\,z\right )$ are the usual right-handed Cartesian coordinates
with the $z$-axis oriented along the symmetry axis of the disk and the Sun located 
on the $x$-axis, $R=\left
(x^2 + y^2\right )^{1/2}$, and $r = \left (R^2 + z^2\right )^{1/2}$.
We also define a triaxial coordinate system for the halo with
coordinates $\mathbf{r}_{t}=\left (x_t,\,y_t,\,z_t\right )$.
Following LMJ09, we fix the $z_t$-axis to be along the symmetry axis
of the disk and write $\mathbf{r}_{t}=
\mathbf{R\Lambda}_\Phi\mathbf{r}$ where
$\mathbf{\Lambda}_\Phi={\rm diag}\left
(1,A^{-1}_\Phi,B^{-1}_\Phi\right )$ and
\begin{equation}
 \mathbf{R}=\left(
\begin{tabular}{ccc}
$\cos \theta$ & $\sin \theta$ & 0 \\
$-\sin \theta$ & $\cos \theta$ & 0 \\
0 & 0 & 1
\end{tabular}
\right)~.
\end{equation}
In this section, we follow LMJ09 (see also \citet{Johnston1999} and
\citet*{Law2005}) and fix the structural parameters of the disk, bulge, 
and halo potentials to the following values:
$M_{b}=3.43\times10^{10}M_{\odot}$, $c=0.7 \textrm{ kpc}$,
$M_{d}=1.0\times10^{11}M_{\odot}$, $a=6.5 \textrm{ kpc}$, $b=0.26
\textrm{ kpc}$, and $d=12 \textrm{kpc}$.  The halo scale velocity,
$\sigma_{h}$, is adjusted so that the local circular speed is
$220~\kms$.  We refer to this model as the LJM model.

\subsection{Observational constraints from the Sgr dSph and stream}

The Sgr system roughly lies in a single plane, which has been
identified as the orbital plane of the Sgr dSph.  To a good
approximation, this plane contains the Sun and is defined by the
orbital pole $\left (l_p,\,b_p\right ) = (273.8^\circ,-14.5^\circ)$
(LM10 and \citet{Johnston2005}).  The Sgr dSph itself is located at
$\left (l_{\rm Sgr},\,b_{\rm Sgr}\right ) = (5.6^\circ,-14.2^\circ)$.

\citet{Ibata1997} found that the stars observed in the central regions
of the Sgr dSph have a mean radial velocity of $171\pm 1\,\kms$, which
is nearly the same as the radial velocity of M54, a globular cluster
usually identified with the center of the Sgr dSph.  Note that this
velocity has been converted to the Galactic Standard of Rest (GSR), a
reference frame centered on the Sun but at rest relative to the
Galactic center.  The conversion from heliocentric velocities to the
GSR depends on the circular speed at the position of the Sun and the
Sun's peculiar velocity $\left (U_\odot,\,V_\odot,\,W_\odot\right )$.
In \citet{Ibata1997} as well as LMJ09 and LM10, the circular speed of
the Sun is assumed to be $220\,\kms$.  We return to this assumption in
Section 3.  The heliocentric proper motion of the Sgr dSph has been
measured by \citet*{Pryor2010} using archival Hubble Space Telescope
(HST) data as $\left (\mu_l,\,\mu_b\right ) = \left (-2.615\pm
0.22,\,1.87\pm 0.19\right ) {\rm mas\,yr}^{-1}$. The heliocentric
distance to the Sgr dSph $D_{\rm Sgr}$ has been estimated by a various
methods and is found to lie in the range from $22-28.4\,{\rm kpc}$
(see Table 2 of \citet{Kunder2009}).

\subsection{Likelihood function and posterior probability distribution}

For simplicity we use the so-called Sagittarius spherical coordinate
system $\left (d,\lambda,\,\beta\right )$ defined by
\citet{Majewski2003} where $d$ is the heliocentric radial coordinate
and $\lambda$ and $\beta$ are angular coordinates. The Sgr dSph is
located at $\lambda=\beta=0^\circ$ while the stream approximately
follows the $\beta=0^\circ$ great circle with $\lambda$ increasing
along the trailing portion of the stream and $\beta$ increasing toward
the orbital pole.  Following LMJ09, we assume that the Sgr dSph can be
modeled as a point mass and that the leading and trailing portions of
the stream follow, respectively, the future and past segments of its
orbit. The model orbit is then compared with the radial velocity and
angular position measurements of M giants from \citet{Majewski2003}
and the angular position measurements from the SDSS study of
\citet{Belokurov2006}.  This method ignores the internal structure and
dynamical evolution of the progenitor and stream.  N-body simulations
allow one to explore these effects but their use is unfeasible in the
present analysis where 500K likelihood calls are required.  (See
however \citet{Varghese2011} who discuss a promising approach in which
a family of kinematic orbits are used to model the ``thick'' tidal
stream.)

For the ``i'th'' data point, we determine the position along the orbit
with the same $\lambda_i$ and, in so doing, arrive at model
predictions for the line-of-sight velocity $v_M\left (\lambda_i\right
)$ and the angular position $\beta_M(\lambda_i)$.  The likelihood
function, or equivalently, the probability of the
data given the model $M$, is then
\begin{eqnarray}
\label{Eq:Like}
p\left (D|M\right )
&=&
\prod_{i=1}^{N_\beta} 
\frac{1}{\left (2\pi\right )^{1/2}\sigma_{\beta,i}}
\exp{\left [-
\frac{\left (\beta_{M}\left (\lambda_i\right )-\beta_i\right )^2}
{2\sigma_{\beta,i}^2}\right ]}
  \nonumber \\
 &\times&\prod_{i=1}^{N_v}
\frac{1}{\left (2\pi\right )^{1/2}\sigma_{v,i}}
\exp{\left [-
\frac{\left (v_{M}\left (\lambda_i\right )-v_i\right )^2}
{2\sigma_{v,i}^2}\right ]}.
\end{eqnarray}
The first term on the right-hand side involves a product over all M
giants in the \citet{Majewski2003} data set as well as the SDSS Sgr
stream fields from \citet{Belokurov2006} whereas the second term
involves only the M giants.  The parameters $\sigma_\beta$ are meant
to account for both observational uncertainties and the intrinsic
thickness of the Sgr stream.  Likewise, the $\sigma_v$ are meant to
account for observational uncertainties in the line-of-sight velocity
and the intrinsic dispersion.

Our aim is to calculate the posterior probability function (PDF) of
the model, which is given by Baye's theorem,
\begin{equation}
 p(M|D,I)=\frac{p(M|I)p(D|M)}{p(D|I)}~,
\end{equation}
where $I$ represents prior information and $P(M|I)$ is the prior
probability on the model M.  The term $p(D|I)$, often referred to as the
evidence, is essentially a normalization factor and does not enter
our calculations.

In general, $M$ is specified by $N_P$ parameters and therefore
$P(M|D,I)$ is an $N_P$-dimensional function.  In order to efficiently
map this multi-dimensional function we use the Stretch-Move (SM) MCMC
algorithm from \citet{Goodman2010}.  All MCMC algorithms require a
proposal distribution or sampler to generate the chain of points in
parameter space.  If the different model parameters have different
scales or are strongly correlated, then it can be extremely difficult to
choose an effective sampler.  And a poorly chosen sampler leads to
poor convergence of the chain.  The SM-MCMC algorithm employs an
ensemble of $N_W$ ``walkers'' where $N_W > N_P$.  The proposal
distribution for a given walker is determined by the other walkers
(rather than by the user).  By design, the algorithm is invariant
under linear re-parameterizations of the model parameters (i.e.,
affine invariant) and the difficulties of choosing a sampler are
avoided.

\subsection{LMJ09 revisited}

We next reexamine the \citet{Majewski2003} and \citet{Belokurov2006}
data sets within the context of the Galactic model described above and
model assumptions that are similar to those used in LMJ09.  The free
parameters are the axis ratios $A_{\Phi}$ and $B_{\Phi}$, the angular position
$\theta$ of the $x_t$-axis in the $x_t-y_t$ plane and the two proper
motion components of the Sgr dSph.  Following LMJ09, we fix $D_{\rm
  Sgr}$ and line-of-sight velocity for the Sgr dSph to be $28\,{\rm
  kpc}$ and $171\,\kms$ (GSR), respectively.  We also follow LMJ09 and
set $\sigma_{\beta,i} = 1.9^\circ$ and $\sigma_{v,i}=12\,\kms$.

We assume uniform priors in $\log(A)$ and $\log(B)$ since the axis
ratios $1:0.5$ and $1:2$ are effectively equivalent.  To avoid
degeneracies, we restrict $A$ to be greater than unity.  The prior for
$\theta$ is assumed to be uniform between $0^\circ$ and $180^\circ$.
Finally, we assume that the prior probability on the components of the
proper motion are uniform across a range that includes both the
best-fit values from LM10 and values observed by \citet{Pryor2010}.

We fit the stream using our SM-MCMC algorithm with 100 walkers and
5000 steps.  Here and throughout this work we discard the first 1000
steps, i.e., the burn-in segment of the chain.  The marginalized PDFs
for $A_\Phi$, $B_\Phi$, and $\theta$ in this run (referred to as the triaxial potential or TP run)
are shown in Figure 1 while a
summary is provided in Table 1.  Interestingly, the most likely shape
from the Bayesian analysis is closer to the results of LM10 than the
results from LMJ09.  The implication is that the parameter search algorithm is 
at least as important as whether the Sgr dSph is modeled as an N-body system 
or a point particle.

\begin{table*}
\centering
\begin{minipage}{126mm}
\centering
 \begin{tabular}{|c|c|c|c|c|c|}
\hline  & $A_\Phi$ & $B_\Phi$ & $A_\rho$ & $B_\rho$ & $\theta$ \\
\hline
\hline
LMJ09 & 1.5 & 1.25 & (2.5) & (1.75) & $0^\circ$ \\
LM10 & 1.4 & 1.4 & (2.2) & (2.2) & $-7^\circ$ \\
TP & 1.49 & 1.34 & (2.48) & (2.03) & $-9^\circ$ \\
TD & (1.49) & (1.38) & 2.48 & 2.15 & $-11^\circ$ \\
\hline
 \end{tabular}
\caption{Axis ratios and angle $\theta$ for the LMJ09, LM10 analyses
  and TP and TD runs as described in Section 2.  Parenthesis denote axis
  ratios that are calculated using the expression $A_\rho = 3\left
  (A_\Phi - 1\right ) + 1$ or its inverse.}
\end{minipage}
\end{table*}

\subsection{Triaxial density}

By design, isopotential surfaces in the halo model described by
Equation \ref{eq:LJMHaloPot} are triaxial whereas the isodensity
surfaces are not.  Indeed, when $A$ and $B$ differ significantly from
$1$, as in the best-fit models of LMJ09 and LM10, isodensity surfaces
become peanut-shaped (see Fig. 2.9 of \citet{BT2008}).  In
simulations, the shape of dark matter halos is inferred by measuring
the axis ratios of isodensity surfaces assuming that these are the
surfaces that are triaxial.  With these issues in mind, we consider an
alternative model in which the halo density is given by
\begin{equation}\label{eq:LJMHaloDens}
\rho_h(R,z) = \rho_h\left (r_t^2\right )
\frac{\sigma_h^2}{2\pi G} \frac{r^2 + 3d^2}{\left (r_t^2 + d^2\right  )^2}
\end{equation}
where $\mathbf{r_t}\equiv \mathbf{R\Lambda_\rho r}$ with
  $\mathbf{\Lambda_\rho} = {\rm diag}\left
    (1,\,A^{-1}_\rho,\,B^{-1}_\rho\right )$.  Note that in the
    spherical limit, $r=r_t$ and Equations \ref{eq:LJMHaloDens} and
    \ref{eq:LJMHaloPot} describe equivalent models.

The force and potential are calculated from Equation
\ref{eq:LJMHaloDens} via the homeoid theorem \citep{BT2008}.  The
potential for any triaxial density is given by
\begin{equation}\label{Eq:homeoid}
\Phi(\mathbf{x})=-\pi G \frac{a_{2}a_{3}}{a_{1}}\int_{0}^{\infty}
d\tau \frac{\psi(\infty)-\psi(m)}
{\sqrt{(\tau+a_{1}^{2})(\tau+a_{2}^{2})(\tau+a_{3}^{2})}}~,
\end{equation}
where the $a_{i}$ are the axis ratios, 
\begin{equation}\label{Eq:M}
m^2=a_{1}^{2}\sum_{i=1}^{3}\frac{x_{i}^{2}}{a_{i}^{2}+\tau}
\end{equation}
is similar to the square of the ellipsoidal radius, and
\begin{equation}
\psi(m)=\int_{0}^{m^{2}}dm^{2}\rho(m^{2})~,
\end{equation}
is an auxiliary function.  In our models $a_{1}=1$, $a_{2}=A_\rho$,
and $a_{3}=B_\rho$.

We consider a Galactic model in which Equation \ref{eq:LJMHaloDens}
rather than Equation \ref{eq:LJMHaloPot} describes the halo.  We also
allow $D_{\rm Sgr}$ to vary within the range of measured values compiled by
\citet{Kunder2009}.  To be precise, we assume a
uniform prior for $D_{\rm Sgr}$ between $22\,{\rm kpc}$ and
$28.5\,{\rm kpc}$.  \citet{Law2005} found that the effect of varying
$D_{\rm Sgr}$ was degenerate with other parameters, particularly the
distance $\Ro$ of the Sun to the Galactic center.  We therefore include
$\Ro$ as a model parameter and assume a Gaussian prior centered on
$8.2\,{\rm kpc}$ with 1-$\sigma$ width of $0.4\,{\rm kpc}$
\citep*{Bovy2009}.  We also include a prior probability for the proper
motion based on the HST analysis of \citet{Pryor2010}.  Finally, we
replace $\sigma_{\beta,i}$ and $\sigma_{v,i}$ in Equation
\ref{Eq:Like} with the free parameters $\epsilon_{\beta,j}$ and
$\epsilon_{v}$, where $j=MG$ for the \citet{Majewski2003} M giants and
SDSS for the \citet{Belokurov2006} SDSS fields.  That is, we {\it
  model} the stream thickness in both angular position and
line-of-sight velocity.  To summarize, our model now includes five
additional parameters, $D_{\rm Sgr}$, $\Ro$, $\epsilon_{MG}$,
$\epsilon_{SDSS}$, and $\epsilon_v$.

The marginalized PDFs for $A_\rho$, $B_\rho$, and $\theta$ are shown
in Figure 2 (referred to as the triaxial density or TD run).  
We also compare the best-fit values with those from the
previous analysis and with the results from LMJ09 and LMJ10.  To do
so, we use the relation $A_\rho = 3\left (A_\Phi - 1\right ) + 1$ from
\citet{BT2008}.  Once again, we find that the Sgr stream favours a
strongly oblate-triaxial halo with isodensity axis ratios that are
greater than 2 and with the symmetry axis closely aligned with the
Sun-GC line.  The uncertainty in the axis ratios is greater, by about
a factor of three, for this run.

In Figure 3, we show the marginalized PDF in the $D_{\rm Sgr}-\Ro$
plane.  The constraint on $\Ro$ comes almost entirely from the prior,
while our fit to the stream data favours the upper range of
observed values for the distance to the Sgr dSph.

In Figure 4, we show the marginalized PDF for the components of the
proper motion, $\mu_l$ and $\mu_b$.  Evidently, the PDFs are
inconsistent with measurements by \citet{Pryor2010} at about the
$2\sigma$-level but are consistent with the favoured values obtained
in LMJ09 and LM10.  Tension between the model and observed values for
the proper motion may indicate a problem with the model assumptions
(e.g., interpretation of the streams as direct tracers of the orbit)
though clearly a more refined measurement of $\mu_l$ and $\mu_b$ is
required.

\section{Toward a more realistic Milky Way model}

\subsection{An updated Galactic model}

The model described by Equations 1-3 is attractive in large part because of
its simplicity since the potential, force, and density can all be
expressed in terms of analytic functions.  However, this model 
does not represent our current understanding of the 
azimuthally-averaged structure of spiral galaxies.
For example,
\citet{Andredakis1995} showed that bulges of spiral galaxies could be
best fit by the S\'{e}rsic law.  Furthermore, since the classic paper by
\citet{Freeman1970}, it has become standard practice to model the
luminosity profile of disks as an exponential.  Finally, the
spherically-averaged density profile of dark matter halos is described
extremely well by the Einasto profile \citep{Merritt2006}.  Therefor we
consider a Milky Way model with these three components.  We assume
that the bulge and disk have constant (but different) mass-to-light
ratios.  The density that produces a S\'{e}rsic surface brightness
profile is \citep{Prugniel1997,Terzic2005}
\begin{equation}\label{eq:Sersic}
\rho_{b}(r) = \rho_{b0}\left(\frac{r}{R_{e}}
\right)^{-p}e^{-b(r/Re)^{1/n}}~,
\end{equation}
where $p = 1 - 0.6097/n + 0.05563/n^2$, $n$ is the S\'{e}rsic index
and $b=b(n)$ is chosen so that the radius $R_e$ encloses half the
total projected mass.  For simplicity, we parameterize the bulge by
the scale velocity
\begin{equation}
\sigma_b\equiv \left (4\pi nb^{n\left (p-2\right )}\Gamma\left (n\left
     (2-p\right )\right )R_e^2\rho_{b0}\right )^{1/2}
\end{equation}
rather than $\rho_{b0}$.  

We assume that the disk density is exponential in the radial direction
(the Freeman Law) and has a sech$^2$ structure in the vertical direction:
\begin{equation}\label{eq:diskdensity}
\rho_d\left (R,\,z\right ) =~\frac{M_d}{4\pi R_d^2z_d}   
\, e^{-R/R_d}\, {\rm sech}^2\left (z/z_d\right )
\end{equation}  
where $M_d$, $R_d$, and $z_d$ are the mass, radial scale length, and
vertical scale height for the disk, respectively.
The disk potential is calculated using the technique of
\citet{Kuijken1995}.  An analytic 'fake' disk density-potential pair,
$\left (\rho_{fd},\,\Phi_{fd}\right )$ is constructed so that
$\rho_{d}=\rho_{fd}+\rho_{r}$ and $\Phi_d = \Phi_{fd} + \Phi_r$ where
$\left (\rho_r,\,\Phi_r\right )$ is the density-potential pair of the
residual.  The fake disk is designed to account for the high-order
moments of the disk potential.  The full potential is calculated by
solving the Poisson equation using a spherical harmonics expansion and
iterative scheme.

The density distribution for our halo model is given by
\begin{equation}
 \rho_{h}(r_{t})=\rho_{0}e^{-\frac{2}{\alpha}((r_{t}/r_{h})^{\alpha}-1)}
\end{equation}
where $r_{t}$ is the triaxial radius, $\rho_{0}$ is a scale density,
$r_{h}$ is the scale radius, and $\alpha$ controls the logarithmic
slope of the density profile.  The potential for this
triaxial density is calculated via Equation \ref{Eq:homeoid}.

To be sure this model represents an approximation to the mass
distribution and gravitational potential of the Galaxy.  The model
disk and bulge are axisymmetric whereas the actual bulge is triaxial
and the disk contains a bar.  It is reasonable to ask whether these
non-axisymmetric structures could mimic the effects of a triaxial
halo.  We note that at pericenter the Sgr dSph is $\sim20$ kpc from
the Galactic center whereas the bulge triaxiality and the bar are
properties of the inner few kpc.  At pericenter, the disk and bulge
contribute $\sim25\%$ and $10\%$ respectively to the gravitational
force.  Furthermore, the components of the potential associated with
the bar and bulge triaxiality are quadruple (or higher order) terms,
which fall off significantly faster than the leading monopole term.
We therefore conclude that the triaxiality of the bulge and the
presence of a bar will have little effect on the orbit of the Sgr dSph
and its stream.

\subsection{Observational constraints}

Our analysis combines kinematic and photometric observations that
constrain the structure of the Milky Way across five orders of
magnitude in radius.  Apart from the constraints due to the Sgr
stream, the observations we consider are similar to those found in
\citep*{Dehnen1998,Widrow2005,Widrow2008} with some notable exceptions.
The starting point is a generalization of
Equation \ref{Eq:Like} to
\begin{equation}\label{Eq:FullLike}
 p(D|\mathbf{\Theta})=\prod_{i=1}^{N_{\rm{obs}}}\prod_{j}^{N_{i}}
\frac{1}{2\pi\sigma_{i,j}}e^{\frac{(M_{i,j}-O_{i,j})^{2}}{2\sigma_{i,j}^{2}}}~,
\end{equation}
where the index $i$ runs over the different types of observational
constraints (e.g. stream angular position and radial velocity) while
the index $j$ runs over individual data points of a particular type of
observation.  The model prediction is $M$, the observation, $O$, and
is the associated error, $\sigma$.

The stream constraints are as before with one proviso: Since we are
treating the Sun's motion relative to the Galactic centre as a free
parameter, we cannot use the GSR radial velocities for the M giants as
published in \citet{Majewski2003} but must use the (observed)
heliocentric velocities and convert to the GSR using the model values
for $v_c(\Ro)$, $U_\odot$, $V_\odot$, and $W_\odot$.

For the innermost region of the Galaxy, we use the compilation of
bulge LOSVD measurements found in
\citet{Tremaine2002}, with the restriction that $r\ge4 ~\rm{pc}$ to
avoid complications from the central black hole.  In addition, we adjust
the dispersion downwards by a factor of 1.07 to account for the
non-spherical nature of the bulge.  We also use the surface brightness
profile at infrared wavelengths from the COBE-DIRBE observations
\citep*{Binney1997}.  These observations extend from $-40^\circ$ to
$40^\circ$ in Galactic longitude and allow us to probe the structure
of both the bulge and disk.  Of course, our model must now include
mass-to-light ratios for the bulge and disk.

We include a number of ``local'' or solar neighborhood constraints in
our analysis.  Recently \citet{Reid2009} used Very Long Baseline radio
observations to determine trigonometric parallaxes and proper motions
for masers throughout the Milky Way's disk.  These measurements were
then used to infer a circular speed at the position of the Sun of
$254\pm 16\,\kms$, a value 15\% higher than the one assumed in LMJ09
and LM10.  The \citet{Reid2009} results were scrutinized by
\citet{Bovy2009} who carried out a Bayesian analysis of the maser data
and also incorporated proper motion measurements of Sgr A$^{*}$.  For
our analysis, we adopt the \citet{Bovy2009} value: $v_c\left
(\Ro\right ) = 244\pm\,13\,\kms$.  

The shape of the circular speed curve in the solar neighborhood is
described by the Oort constants.  We adopt the constraints
$A=14.8\pm0.8~\kms \textrm{kpc}^{-1}$ and $B=-12.4\pm0.6~\kms
\textrm{kpc}^{-1}$ from \citet{Feast1997}.  We use the disk surface
density measurement of $\Sigma_{d}=49\pm9~\msol~\rm{pc}^{-3}$ from
\citet{Flynn1994} and the vertical force measurement of
$|K_{z}(1.1~\textrm{kpc})| = \left ({2\pi G}\right )71\pm 6 ~\msol
\textrm{pc}^{-2}~,$ from \citet{Kuijken1991}.  These values are in
good agreement with similar studies (see, for example,
\citet{Holmberg2004}).

Observations of the Galactic circular speed curve are typically
divided into measurements inside and outside the solar circle.  The
inner rotation curve is usually presented in terms of the terminal
velocity, which is defined as the peak velocity along a particular
line-of-sight defined by the Galactic coordinates $b=0$ and $l$ where
$|l|<\pi/2$.  If we assume that the Galaxy is axisymmetric, then
$v_{\rm term} = v_c\left (R\right ) - v_c\left (\Ro\right )\sin{l}$.
We use data from \citet{Malhotra1995} with the restriction that $\sin
l\ge 0.3$ so as to avoid distortions due to the bar.  The outer
rotation curve requires a more in depth discussion.  The circular
speed $v_c(R)$ is related to the observed line-of-sight velocity
$v_{lsr}$ of a kinematic tracer in the disk through the expression
\begin{equation}
W(R)=\frac{\Ro}{R}\,v_{c}(R)-v_{c}(\Ro)=\frac{v_{lsr}}{\cos b \cos
  l}~.
\end{equation}
We use observations of HII regions and reflection nebulae from
\citet{Brand1993} and Carbon stars from \citet{Demers2007} with the
restriction that $l\le 155^\circ$ or $l\ge 205^\circ$, $d\ge 1 $ kpc,
and $W\le 0$ so as to avoid complications due to non-circular motions.
The errors in the measurements are propagated to errors in $W$.
Additionally, 'noise' parameters for $d$ and $v_{lsr}$ are added in
quadrature to the quoted errors to account for the potential for
unknown systematic errors.  That is, the velocity error for the $j$'th
data point is $\sigma_{v}^{2}=\sigma^{2}_{v,m}+\epsilon^{2}_{u}$ where
$\sigma_{v,m}$ is the measured error (the $i$ subscript has been
suppressed) and $\epsilon_{u}$ is the noise parameter.  Similarly, we
have $\sigma_{d}^{2} = \sigma^{2}_{d,m} + d^2\epsilon^{2}_{d}$.  Note
that we allow for different noise parameters for the \citet{Brand1993}
and \citet{Demers2007} data sets.

Finally, we follow \citet{Dehnen1998} and use $ M(r< 100 \textrm{
  kpc})=(7\pm2.5)\times10^{11}\msol$ as a constraint on the mass of
the Milky Way at large Galactocentric radii.  This constraint is based
on the study of satellite kinematics by \citet{Kochanek1996}, the locally
determined 'escape' speed, and the timing of the local group.  It is
also based on modelling of the Magellanic Clouds and stream by
\citet{Lin1995}.  The large error in $M\left (r<100\,{\rm kpc}\right
)$ reflects the potential for large systematic errors.  In principle,
the Sgr system should be able to constrain the mass distribution of
the Milky Way in the $30-100\,{\rm kpc}$ range.

\section{Results}

We now present results based on the constraints described in the
preceding Section.  As before we use the SM-MCMC algorithm with 100
walkers and 5000 steps.  Our model is described by twenty-eight free
parameters: nine structural parameters for the disk, bulge and halo,
mass-to-light ratios for the bulge and disk, two axis ratios and an
orientation angle for the triaxial halo, the distance of the Sun to
the Galactic center and its motion relative to the local standard of
rest, the heliocentric distance to the Sgr dSph and its proper motion,
and seven ``noise'' parameters.  The initial angular position and
heliocentric velocity of the Sgr dSph are held fixed.  Our priors for
the disk scale length and scale height are based on \citet{Juric2008}
while our priors for the Sun's peculiar velocity are based on values
from \citet{BM1998}.  

In Table 2 we present a summary of the statistics for the model parameters.
In particular, we give the mean and variance of the marginalized PDF for each 
parameter.  For convenience, we also list the prior used for each parameter.

\begin{table*}
\begin{minipage}{126mm}
\centering
 \begin{tabular}{|c|c|c|c|c|c|}
\hline
$\Theta$ & $\langle\Theta\rangle$ & ${\rm Var}\left (\Theta\right )^{1/2}$ & Prior & Lower Limit & Upper Limit \\
\hline
$n$ & 1.77 & 0.37 & uniform & 0.6 & 4.\\
$\sigma_{b}$ & 314 & 23 & log & 100. & 500. \\
$r_{e,b}$ & 0.60 & 0.07 & log & 0.1 & 2. \\
$M/L_{b}$ & 0.69 & 0.20 & uniform & 0.4 & 2. \\
$M/L_{d}$ & 1.09 & 0.08 & uniform & 0.4 & 2. \\
$M_{d}$ & $5.2\times 10^{10}$ & $0.5\times 10^{10}$ & log & $1.8\times10^{10}$ & $7.0\times10^{11}$ \\
$\rho_{0}$ & $1.4\times 10^{-3}$ & $0.4\times 10^{-3}$ & log & $2\times 10^{-3}$ & $7\times 10^{-3}$\\
$r_{e,h}$ & 11.1 & 1.3 & log & 2 & 35 \\
$\alpha$ & 0.23 & 0.07 & uniform & 0 & 0.5\\
$A$ & 3.58 & 0.45 & log & 1 & 5\\
$B$ & 2.58 & 0.25 & log & 0.2 & 5 \\
$\theta$ & -5.4$^\circ$ & $1.4^\circ$ & uniform & $-90^\circ$ & $90^\circ$\\
$d$ & 25. & 1. & uniform & 22 & 28.5\\
$\epsilon_{d,B}$ & 0.42 & 0.05 & log & 0.01 & 0.7 \\
$\epsilon_{v,B}$ & 3.7 & 1.4 & log & 1 & 60\\
$\epsilon_{d,D}$ & 0.49 & 0.03 & log & 0.01 & 0.7 \\
$\epsilon_{v,D}$ & 53. & 14. & log & 1 & 60\\
$\epsilon_{\rm MG}$ & 6.9 & 0.5 & uniform & 1 & 20 \\
$\epsilon_{\rm SDSS}$ & 3.0 & 1.3 & uniform & 1 & 20\\
$\epsilon_{u}$ & 33. & 18. & uniform & 1 & 40\\
\hline
& & & Prior & Width & Mean \\
\hline
$R_{e.d}$ & 2.80 & 0.12 & Gaussian & 1. & 3.5\\
$Z_{e,d}$ & 0.48 & 0.08 & Gaussian & 0.06 & 0.3\\
$U_\odot$ & -9.8 & 0.04 & Gaussian & 0.4 & -10.0 \\
$V_\odot$ & -5.6 & 0.7 & Gaussian & 0.6 & -5.25\\
$W_\odot$ & 7.3 & 0.4 & Gaussian & 0.4 & 7.2\\
$\Ro$ & 8.4 & 0.2 & Gaussian & 0.4 & 8.2\\
$\mu_{l}$ & -2.39 & 0.16 & Gaussian & 0.22 & -2.61\\
$\mu_{b}$ & 1.91 & 0.13 & Gaussian & 0.19 & 1.87\\
\hline
 \end{tabular}
\caption{Statistical summary of results from the MCMC run described in
  Section 4.  Column 1 --- parameter; column 2 --- mean value; column
  3 --- variance; column 4 --- type of prior used.  Columns 5 and 6
  give the lower and upper bounds for prior in the case of either a
  uniform or log prior and the width and mean for the case where the
  prior is a Gaussian.  The units are $\kms$, kpc, $\msol$,
  $\msol~\textrm{pc}^{-3}$, $\mas$ for velocity, distance, mass,
  density, and proper motions respectively.}
  \end{minipage}
\end{table*}

\subsection{General properties of the Galaxy}

Table 3 presents a summary of the statistics for the observables based
on our MCMC analysis.  For comparison, we include the predictions
from the LJM model.  We see that our analysis favors a somewhat lower
value for the circular speed than advocated by \citet{Bovy2009} though
the discrepancy is only at the $1.5\sigma$-level.  More importantly,
our model is consistent with estimates for both the disk surface
density and vertical force in the solar neighborhood in contrast with
the LJM model, which predicts values that are too high, implying that their
disk mass is too large.

\begin{table*}
\begin{minipage}{126mm}
\centering
 \begin{tabular}{|c|c|c|c|c|}
\hline Observation &  Observed & LJM & This Work \\
\hline
$v_{c}$ ($\kms$) & $244\pm 13$ & $220$ & $225\pm 4$ \\
Oort $A$ ($\kmskpc$) & $14\pm 0.8$ & $13.0$ & $14.4\pm 0.3$ \\
Oort $B$ ($\kmskpc$)& $-12.4\pm 0.6$ &$-14.5$& $-12.6\pm 0.4$ \\
$\Sigma_{d}$ ($\msol \textrm{pc}^{-2}$)& $49\pm 9$ & $95$ & $49\pm 4$\\
$\left (2\pi G\right )^{-1}{|K_{z}(1.1~\textrm{kpc})|}$ ($\msol \textrm{pc}^{-2}$)& $71\pm 6$ &$100.1$ & $75\pm 5$ \\
$M(r<100~\textrm{kpc})$ ($10^{11}\msol$)& $7\pm 2.5$ &$7.96$& $6.99\pm 0.63$\\
\hline
 \end{tabular}
\caption{Mean values for observables from the LJM model and this work.}
\end{minipage}
\end{table*}

In Figure \ref{Fig:BulgeDisp} we show the LOSVD measurements from
\citet{Tremaine2002} and the inferred values from the model.  Our
model does a good job in matching the general value for the dispersion
profile though the data show a clear peak at 200 pc and decline for
larger radii whereas the model dispersion profile is relatively flat.
The implication is that the simple S\'{e}rsic-Prugniel \& Simion model
may be inadequate for the inner-most regions of the bulge.  We see
that the LJM model predicts LOSVD's that are 1.5-2 times higher than
the measured velocities, which implies that their bulge mass is 2-4
times to high.

The surface brightness toward the bulge is shown in Figure
\ref{Fig:BulgeSB}.  Both our model and the LJM model do an acceptable
job in fitting the data.  The inner rotation curve is shown in Figure
\ref{Fig:InnerVel} and again, an acceptable fit is obtained.  Note
that the LJM prediction rises more steeply at small $|l|$ as compared
with our model and reaches a peak circular speed that is higher by a
factor of 1.5.

We show the outer rotation curve in Figure \ref{Fig:OuterVel}. In
contrast with the inner regions of the Galaxy, the LJM model and ours
have a very similar shape.  This point is further illustrated in
Figure \ref{Fig:RotCurve} where we show the full circular speed curve
across a wide range in radii.  For comparison, we also show the
\citet{Xue2008} rotation curve, though this data was not used in our
analysis.  The LJM model and ours yield remarkably similar circular
speed curves even though the bulge-disk-halo decomposition of $v_c$ is
very different. We note that the peak contribution to $v_c$ from the 
bulge is nearly twice that for our best-fit model, a result echoed in 
Figure \ref{Fig:BulgeDisp}  The peak contribution to 
$v_c$ from the disk is roughly the same in the
LJM model as in our best-fit model, though in the LJM model, the peak
occurs at R=10 kpc whereas in ours, it occurs at about 5-6 kpc.  Recall
that for an exponential disk, the peak in $v_c$ is at 2.2 disk scale lengths,
so the implication is that the LJM disk has an effective scale length
of ~ 5 kpc, which is roughly a factor of two larger than the standard
values for the Milky Way.

\subsection{The Sgr stream and halo triaxiality}

In Figure \ref{Fig:StreamFit} we show the angular position $\beta$ and
heliocentric velocity $v_r$ of stars that are presumed to be members of the Sgr
stream.  A few comments are in order.  First, the M-giant data is
considerably noisier than the SDSS fields though the latter only
covers a portion of the leading arm of the Sgr stream.  In Table 2, the
stream-thickness (or noise) parameters are $\epsilon_{\rm
  MG}=6.9^\circ$ and $\epsilon_{\rm SDSS}=3^\circ$.  Thus, the
analysis accounts for the intrinsic scatter in the M-giant data by
choosing a larger noise parameter.  This choice has the effect of
reducing the weight the individual M-giant data points as compared
with the SDSS fields in the likelihood calculation.

The PDFs for $A_\rho$, $B_\rho$, and $\theta$ are shown in Figure
\ref{Fig:StreamShape}.  The inferred shape is still roughly oblate
with the short axis nearly aligned with the Sun-GC line.
In fact, the preferred model is more strongly aspherical; the
best-fit values for $A_\rho$ and $B_\rho$ are even larger than were found
in the analysis presented in Section 2.  Note however that the
uncertainties in these values is larger than before.

Regardless of the model or the suite of constraints, the Sagittarius
stream consistently points to a halo model that is oblate-triaxial
with the short axis of the halo approximately coincident with the
Sun-GC line.  
What properties of the Sgr stream drive the fit of
the halo in this direction?  
In Figure \ref{Fig:HaloAlign} we show
the isodensity contours for the total Milky Way model in the $\beta=0$
plane.  Also shown are the principle axes of the halo 
projected on to this plane.  Finally, we show positions
of the M giant and SDSS fields along the stream.  
Roughly speaking, the stream (and presumably, the orbit of the Sgr
dSph) are elongated parallel to the plane of the disk.  In general, if
isopotential surfaces in a particular plane are ellipses, then at
least some of the particle orbits in that plane will be elongated
perpendicular to the long axis of the potential.  Thus, current observations of the Sgr stream
may require a halo that is elongated perpendicular to the plane of the
Galactic disk, as found in LMJ09, LM10 and in the present work.

\subsection{Mass distribution}

The Sgr stream inhabits the region of the Galaxy between 20-60 kpc,
which is roughly where the gravitational potential transitions from
disk-dominated to halo-dominated.  The stream therefore has the
potential to help break the disk-halo degeneracy and constrain the
mass distribution beyond the disk.  To explore this point, we
re-analyze the model without the Sgr stream constraint.  The result
for the cumulative mass profile is shown in Figure
\ref{Fig:MassProfiles}.  We see that uncertainties in the mass at 
$r=100$ kpc are reduced
from 0.3 dex to 0.1 dex by including the Sgr stream,.

As noted above, and illustrated in Figure \ref{Fig:MassProfiles}, the
mass distribution for our model and for the LJM model are very similar
though the mass decomposition are very different.  In particular, the
disk and bulge masses for our model are lower in our models by factor
of 2-4.

\section{Conclusions}

We have performed a Bayesian analysis designed to model Milky Way that
incorporates data for the Sgr stream together with a broader
suite of observational constraints.  Our analysis allows us to not
only infer the shape of the dark matter halo, but also its
spherically-averaged mass distribution.  We find that our Bayesian
approach, which uses single-particle orbits, yields results for the
shape parameters of the halo that are consistent with the N-body-based
approach of LM10.  Moreover, the Sgr stream reduces the
uncertainty in the Milky Way mass profile by a factor of $\sim 2$.

Our analysis has not resolved the central question posed by LMJ09 and
LM10: Why does the Sgr stream appear to favour such a strongly
triaxial halo?  As discussed in the introduction, triaxial halo models
favoured by LMJ09, LM10, and this work are outliers in the
distribution of halo shapes predicted by the standard $\Lambda$-CDM
cosmology.  While the Galactic halo may indeed have this shape, we
caution the reader that these analyses make use of a single stream.
One should be suspicious that the halo model shares the same
approximate orientation as the orbital plane of the Sgr stream.

In principle, the use of multiple stellar streams with different
orbital planes can improve the situation.  Indeed, the GD-1, Orphan,
and Monoceros streams have been already been extensively studied and,
in some cases, used to constrain Milky Way models (see, for example
\citet{Koposov2010,Newberg2010,Penarrubia2005}).  As discussed in
\citet{Koposov2010}, the GD-1 stream is a particular attractive
feature since it is extremely narrow and therefore has a simple
internal structure as compared with the Sgr stream.  However, the GD-1
and Orphan streams cover a significantly smaller angular extent than
the Sgr stream and therefore may have limited value in constraining
the potential.

The tidal debris of satellite galaxies may, in fact, provide other
less direct clues about halo triaxiality.  N-body simulations by
\citet{Rojas-Nino2012} (see also \citet{Penarrubia2009}) suggest that
the morphology of tidal debris depends on the shape of the host
galaxy's halo and in particular, the degree to which it departs from
spherical symmetry.

The results presented in this work and in LMJ09 and LM10 provide a
tantilizing, if not unsettling picture of the Galactic halo.  Clearly,
further analysis and more data are required to establish what the
shape of the dark halo is.  Advances will likely come from a
combination of improved modelling techniques, detailed N-body
simulations and new data from the the next generation of surveys, such
as Gaia \citet{Perryman2002} and LSST \citet{Ivezic2008}.

{We thank K. Johnston, R. Ibata, D. Hogg, and D. Foreman-Mackey for
  insightful comments and suggestions.  LMW acknowledges the financial
  support of the Natural Sciences and Engineering Research Council of
  Canada through the Discovery Grant program.  This research made use
  of the \textbf{Python} programming language and the open-source
  modules \textbf{numpy} and \textbf{matplotlib}.}



\clearpage
\centering
\begin{figure*}
\centering
  \begin{minipage}{126mm}
    \includegraphics[width=126mm]{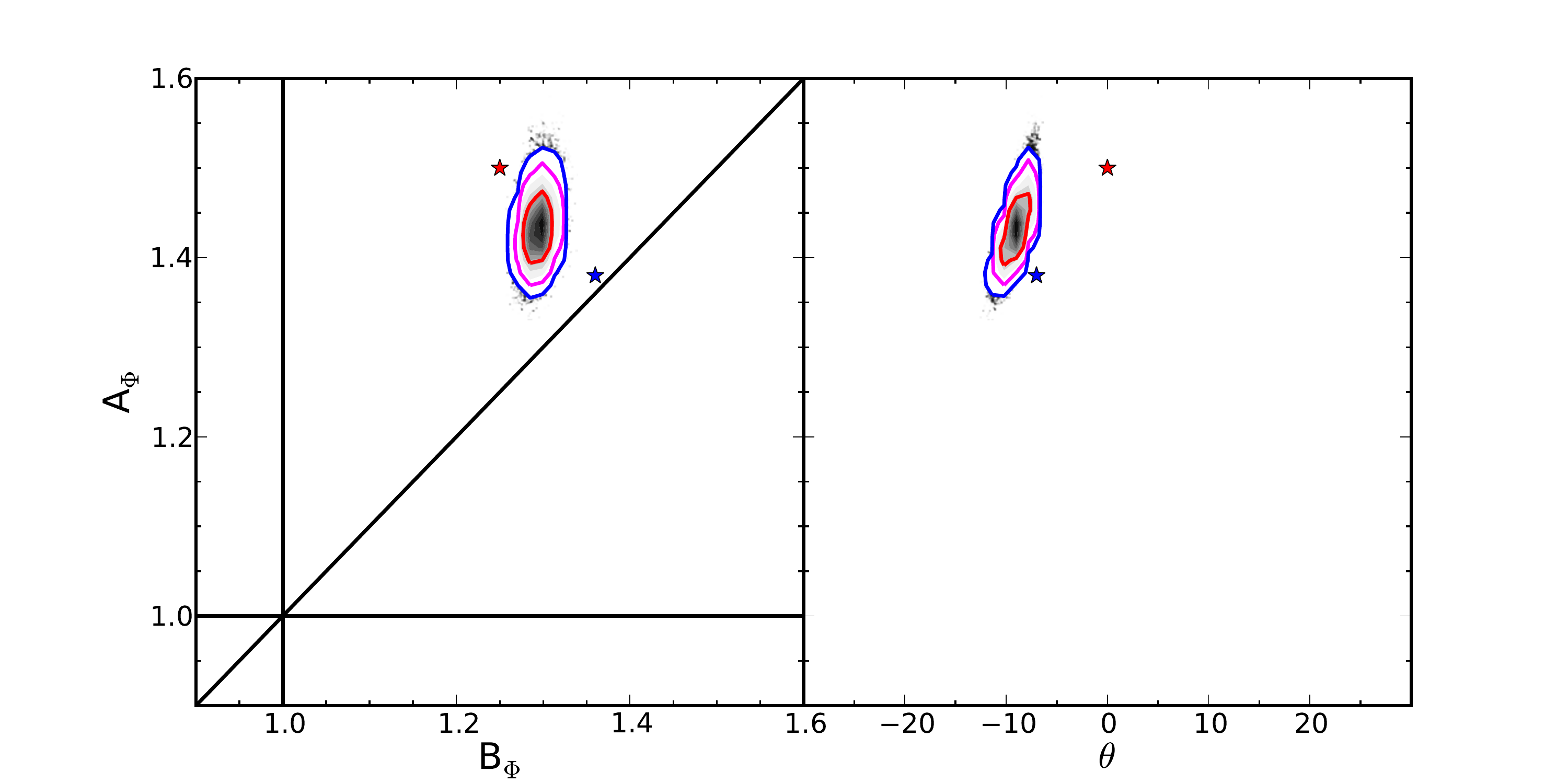}
    \caption{Marginalized PDFs for $A_\Phi$, $B_\Phi$, and $\theta$ from
      the MCMC run described in Section 2.4 (TP).
      Contours enclose 68\%, 95\%, and 99\% of the probability in either
      the $A_\Phi-B_\Phi$ or $A_\Phi-\theta$ planes (left and right
      panels, respectively).  The black points indicate models from the
      MCMC run that lie outside the 99\% contour.  The stars show the
      best-fit parameters from LMJ09 (red) and LM10 (blue).  The diagonal,
      horizontal, and vertical lines in the left-hand panel are for
      reference.  Models along the diagonal line have $A_\Phi=B_\Phi$ and
      are axisymmetric with the symmetry axis coincident with the $x_t$
      axis. Along the upper portion of this line the models are oblate.
      Models along the horizontal $A_\Phi=1$ line are axisymmetric with
      the symmetry axis coincident with the $z$-axis (i.e., the symmetry
      axis of the disk.)}
  \end{minipage}   
\end{figure*}

\begin{figure*}
\centering
  \begin{minipage}{126mm}
    \includegraphics[width=126mm]{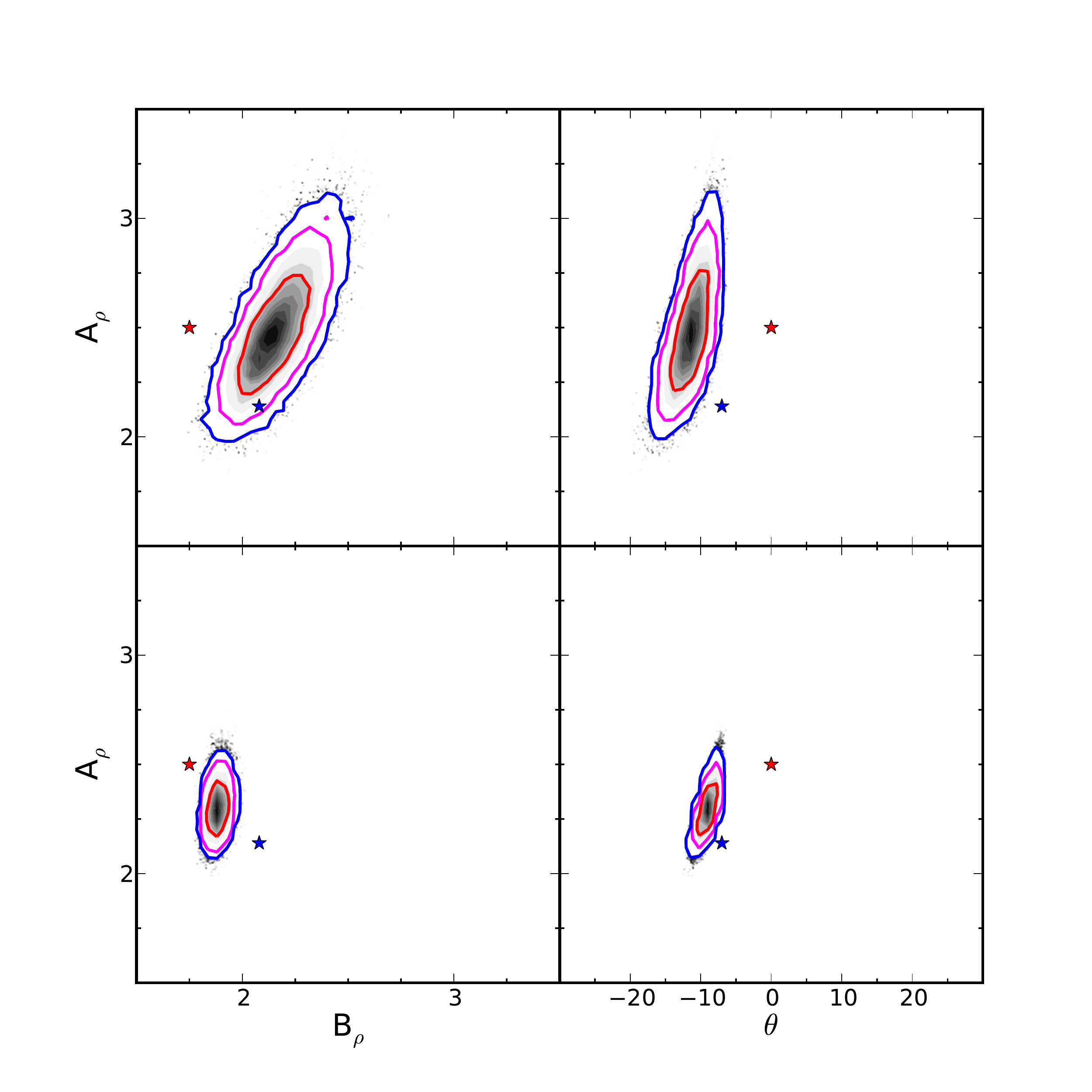} 
\caption{The PDFs for $A_\rho$, $B_\rho$ and $\theta$ for TD as
  described in Section 2.5 (upper panels) and TP (lower panels).
  For the upper panels, $A_\Phi$ is transformed to $A_\rho$ using the
  expression $A_\rho = 3\left (A_\Phi - 1\right ) + 1$ and likewise
  for $B$.  The contours, black points, and symbols are the same as in
  Figure 1.}
  \label{Fig:DensShapePot}
  \end{minipage}
\end{figure*}

\begin{figure*}
\centering
  \begin{minipage}{126mm}
    \includegraphics[width=126mm]{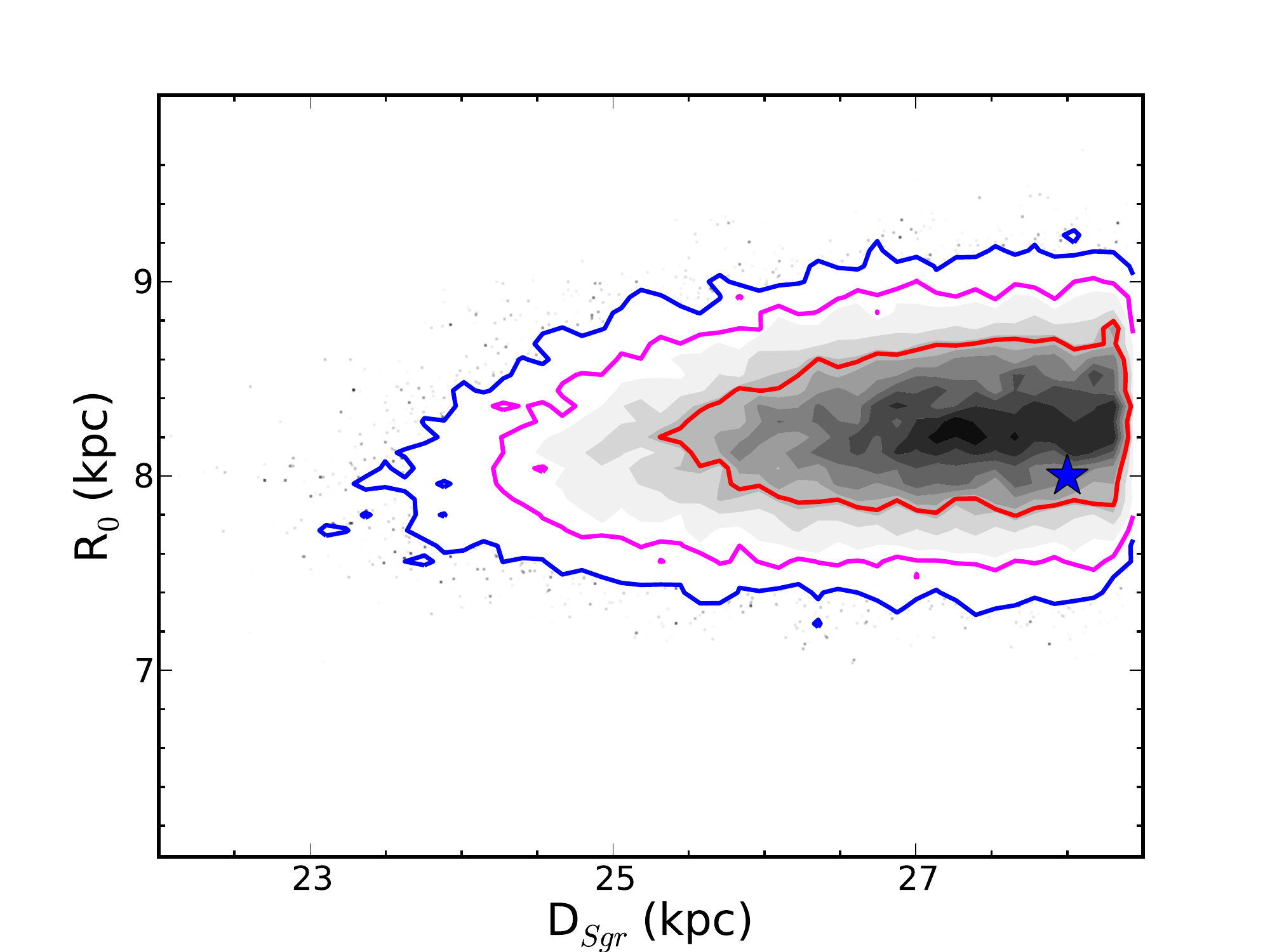} 
\caption{The marginalized PDF in the $\Ro-D_{\rm{Sgr}}$ from TD.
  The contours and black points are the same as Figure 1. The blue
star shows the values assumed by LMJ09 and LM10.}
\label{Fig:DensR0D}
\end{minipage}
\end{figure*}

\begin{figure*}
\centering
  \begin{minipage}{126mm}
    \includegraphics[width=126mm]{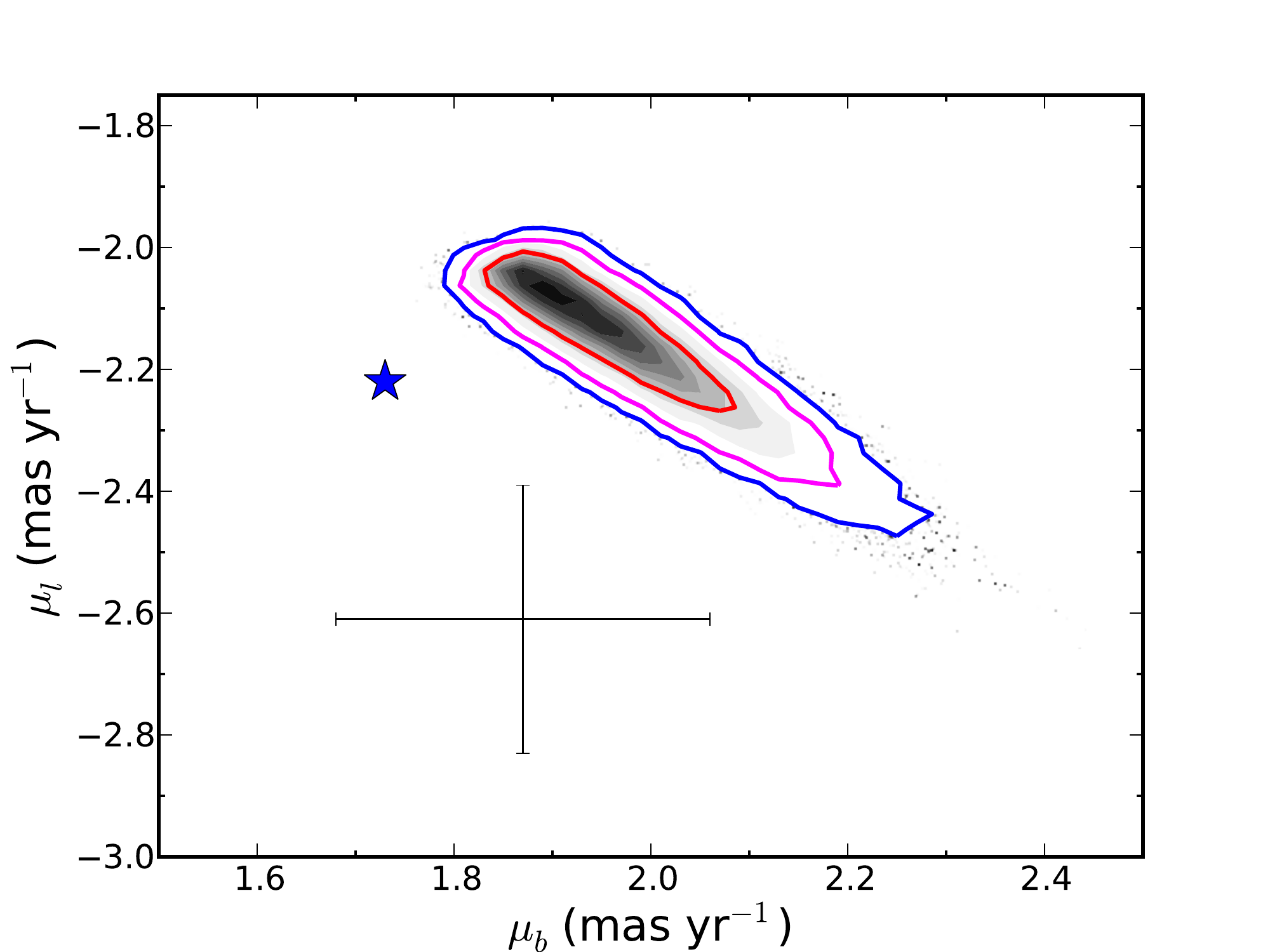} 
\caption{The PDF in the $\mu_l-\mu_b$ plane.  The contours and black
  points are the same as Fig. 1.  The black cross shows the results
  from the \citet{Pryor2010} analysis.  The blue star is the value
  favoured by LM10.}
  \label{Fig:DensPropMot}
  \end{minipage}
\end{figure*}

\begin{figure*}
\centering
  \begin{minipage}{126mm}
    \includegraphics[width=126mm]{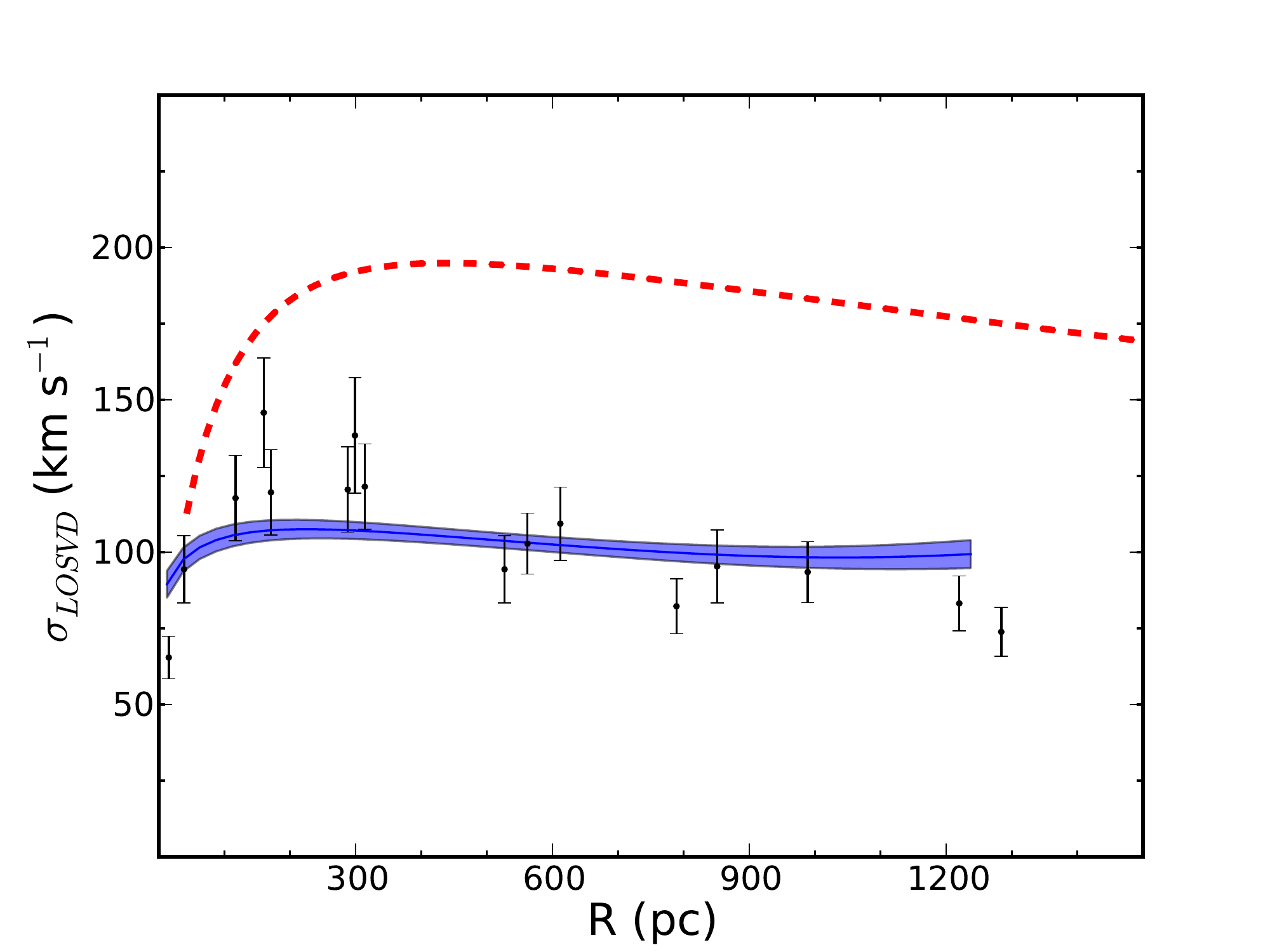}
\caption{Velocity dispersion profile as a function of projected
  radius.  The thin blue line shows the velocity dispersion averaged
  over the chain.  The shaded region gives the 68\% credible region.
  The red dashed curve is the prediction for the LJM model.  The black
  points are from \citet{Tremaine2002}.}
  \label{Fig:BulgeDisp}
  \end{minipage}
\end{figure*}

\begin{figure*}
\centering
  \begin{minipage}{126mm}
    \includegraphics[width=126mm]{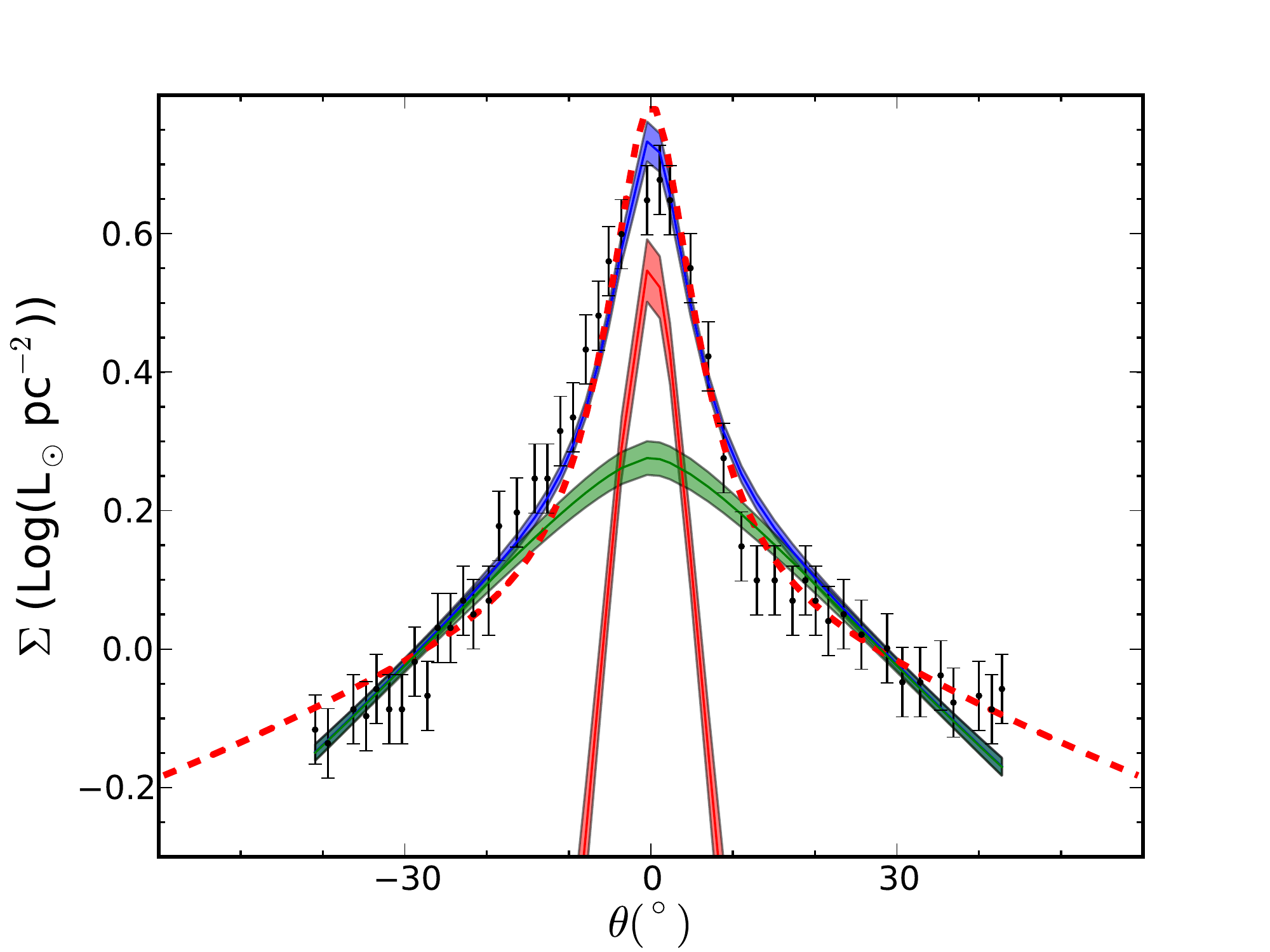}
\caption{Surface brightness as a function $l$.  The thin lines and
  shaded regions are the average and 68\% credible regions for the total
surface brightness (blue), the bulge (red) and the disk (green).
The dashed red curve is prediction for the LJM model.}
\label{Fig:BulgeSB}
\end{minipage}
\end{figure*}

\begin{figure*}
\centering
  \begin{minipage}{126mm}
    \includegraphics[width=126mm]{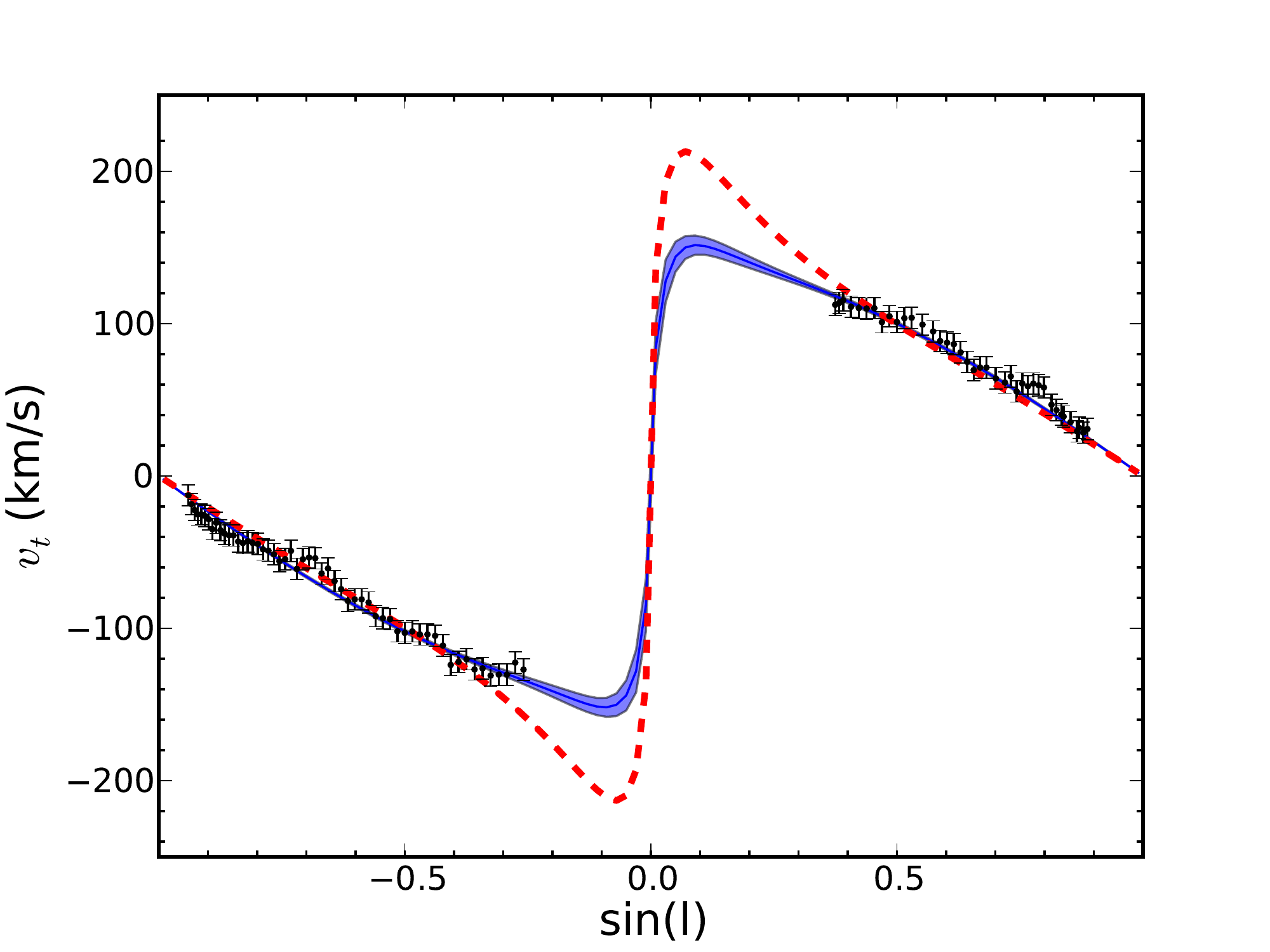}
\caption{Terminal velocity as a function of $\sin{l}$.  The thin blue
  line and shaded region are the average and 68\% credible region,
  respectively while the dashed red curve shows the prediction for the
  LMJ model.  The black points are the data from
  \citet{Malhotra1995}.}
  \label{Fig:InnerVel}
  \end{minipage}
\end{figure*}

\begin{figure*}
\centering
  \begin{minipage}{126mm}
    \includegraphics[width=126mm]{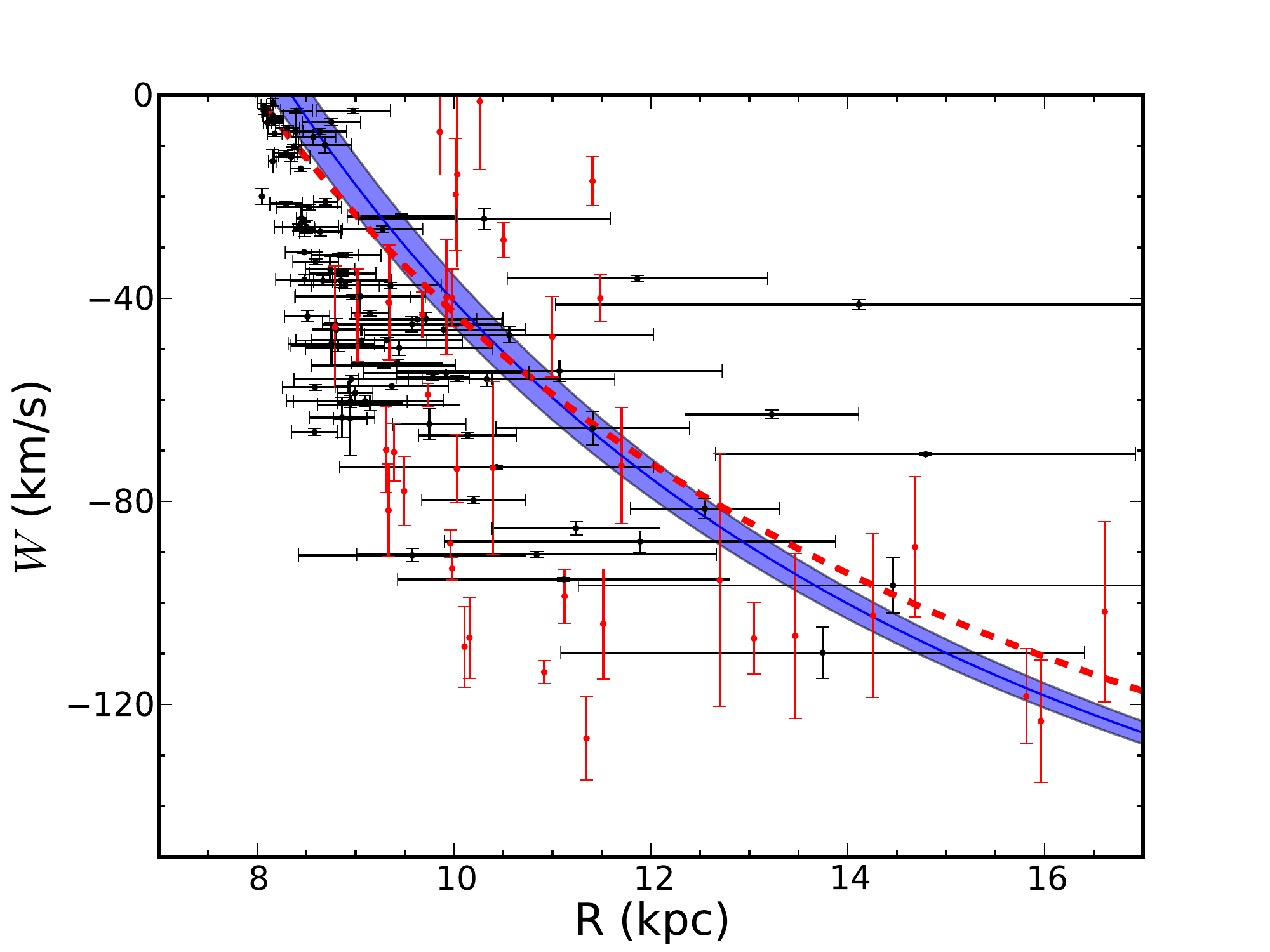}
\caption{Outer rotation curve as a function of projected radius $R$.
  The curves are as in Figure 5.
  The black points are from \citet{Brand1993} and the red points are
  from \citet{Demers2007}.}
  \label{Fig:OuterVel}
  \end{minipage}
\end{figure*}

\begin{figure*}
\centering
  \begin{minipage}{126mm}
    \includegraphics[width=126mm]{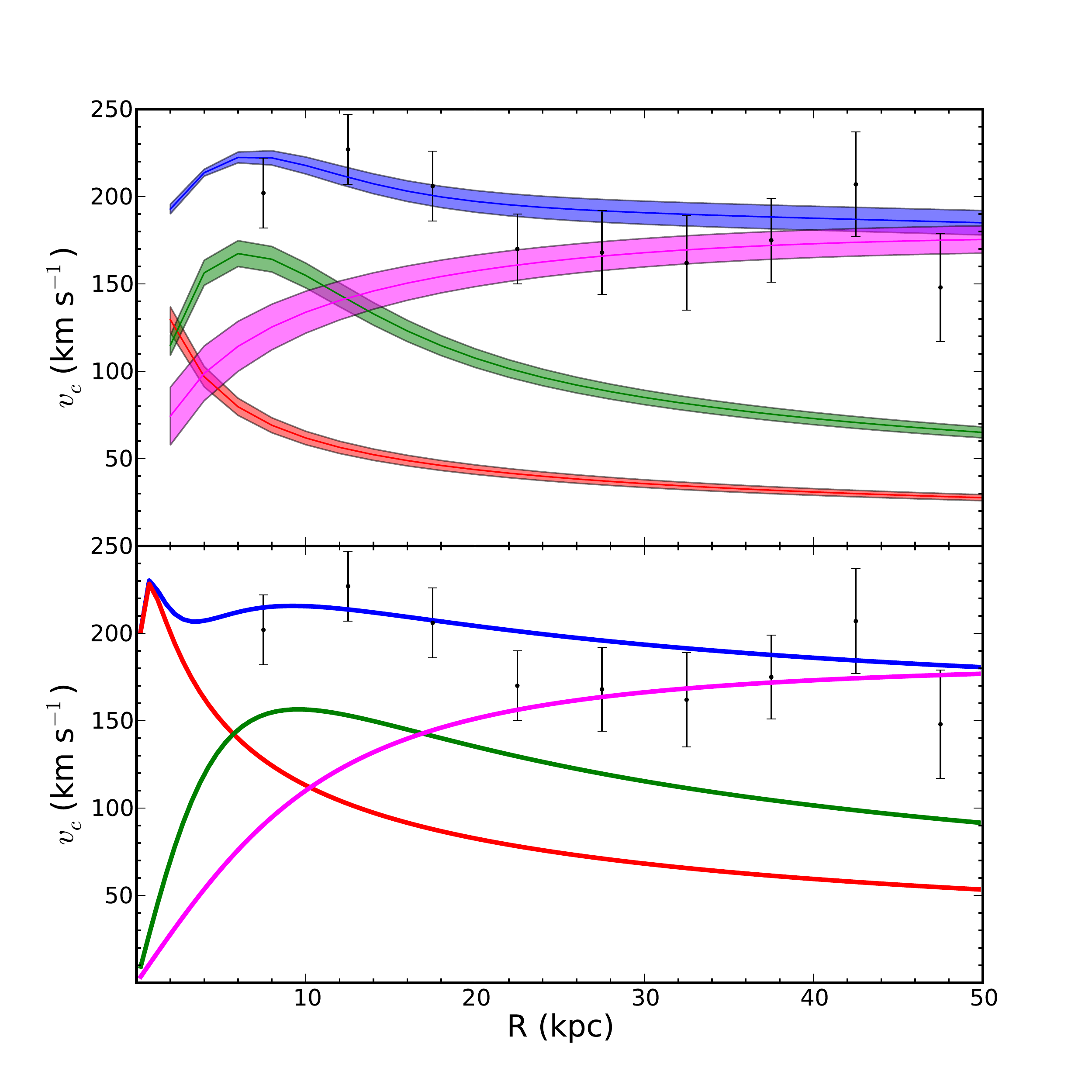} 
\caption{Composite curcular speed curve as a function of radius.
Lines and shaded regions show the average the 68\% credible regions (upper panel)
for the total circular speed curve (blue) and the contributions from
the bulge (red), disk (green), and halo (magenta). The dashed curves (lower panel)
are the prediction from the LJM model.  Black points are from
\citet{Xue2008}. These points are {\it not} used in the fit but
are shown as a consistency check.}
\label{Fig:RotCurve}
\end{minipage}
\end{figure*}

\begin{figure*}
\centering
  \begin{minipage}{126mm}
    \includegraphics[width=126mm]{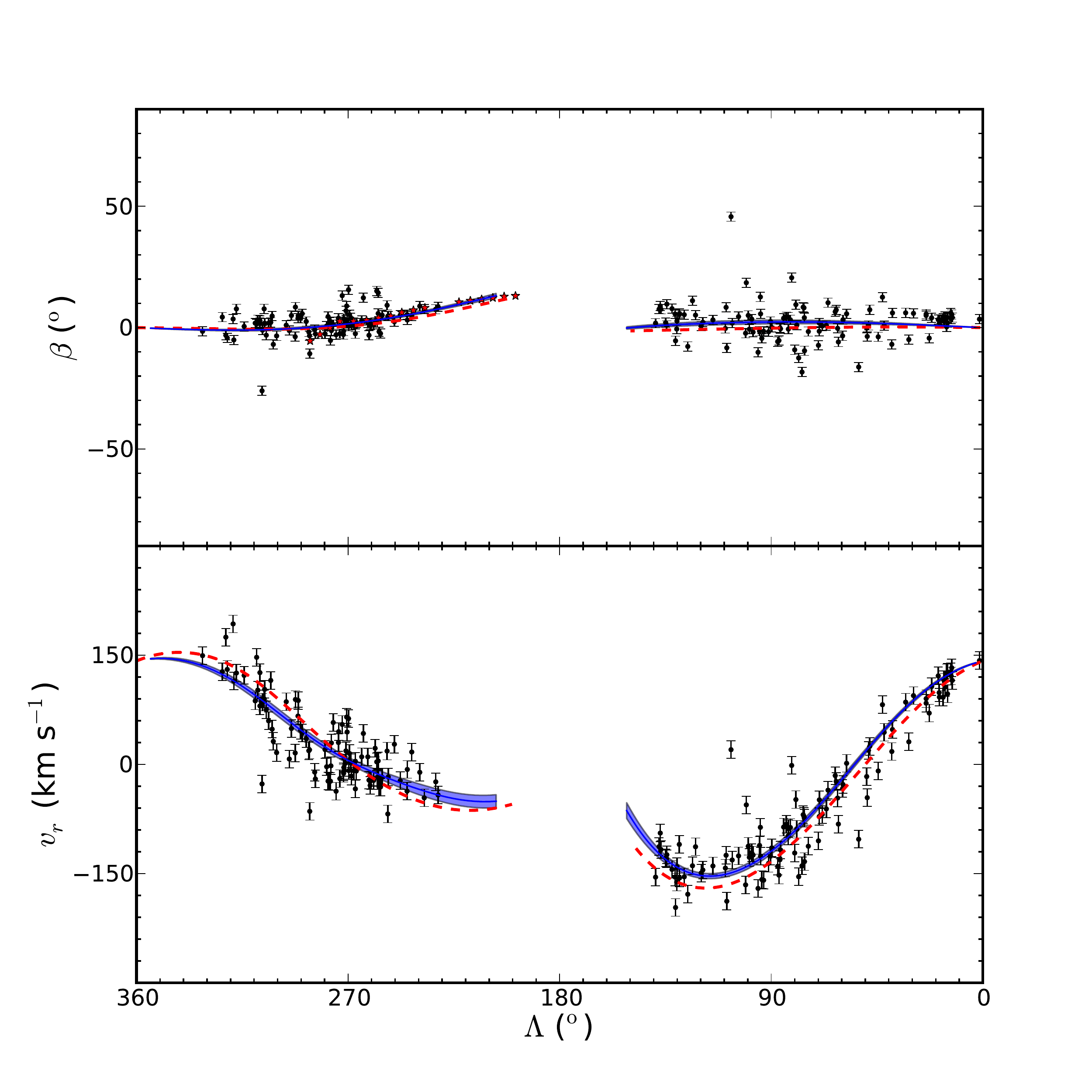}
\caption{Angular postion and line-of-sight velocity as a function of
  position along the stream.  The upper panel shows the angular
  position perpendicular to the stream, $\beta$, as a function of
  angular position along the stream, $\lambda$.  The lower panel shows
  the heliocentric radial velocity, $v_r$, as a function of $\lambda$.
  The thin blue line, shaded region, and dashed curve are as in Figure
  5.  The black points are the 2MASS M giants while the red stars are
  the \citet{Belokurov2006} SDSS fields.  The red dashed line is the
  predictions for the best-fit model of LMJ09.  While the LM10 should be considered
  superior to the LMJ09 model, it produces the fit to the data using an N-body representation
  of the dwarf rather than the single particle approximation.  Since we also use the single
  particle approximation, it is more appropriate to show that comparison.}
  \label{Fig:StreamFit}
  \end{minipage}
\end{figure*}

\begin{figure*}
\centering
  \begin{minipage}{126mm}
    \includegraphics[width=126mm]{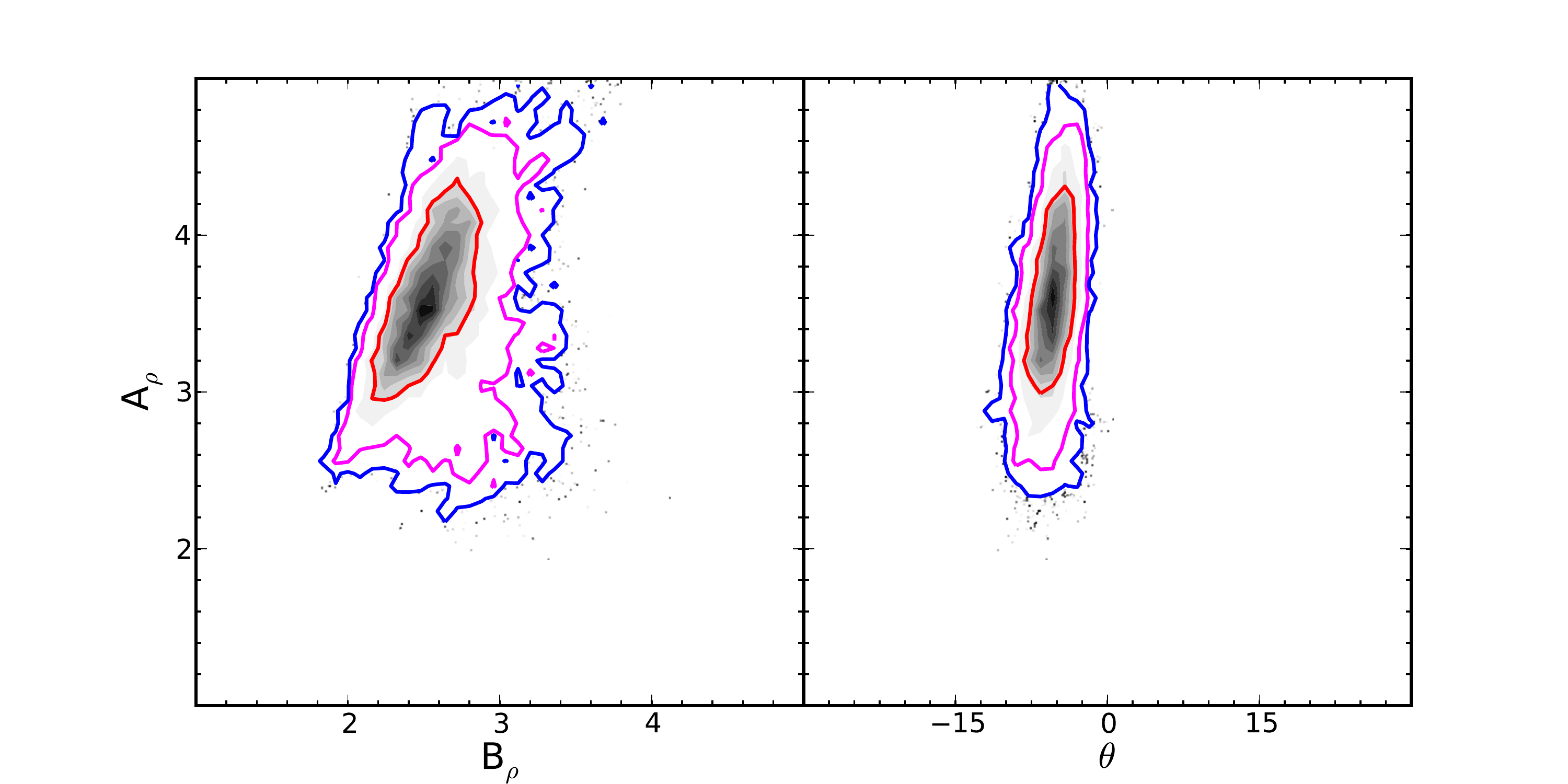} 
\caption{Marginalized PDFs for $A_\rho$, $B_\rho$, and $\theta$ for 
the model introduced in Section 4.  Contours, points and symbols are
the same as those used in Figure 1.}
\label{Fig:StreamShape}
\end{minipage}
\end{figure*}

\begin{figure*}
\centering
  \begin{minipage}{126mm}
    \includegraphics[width=126mm]{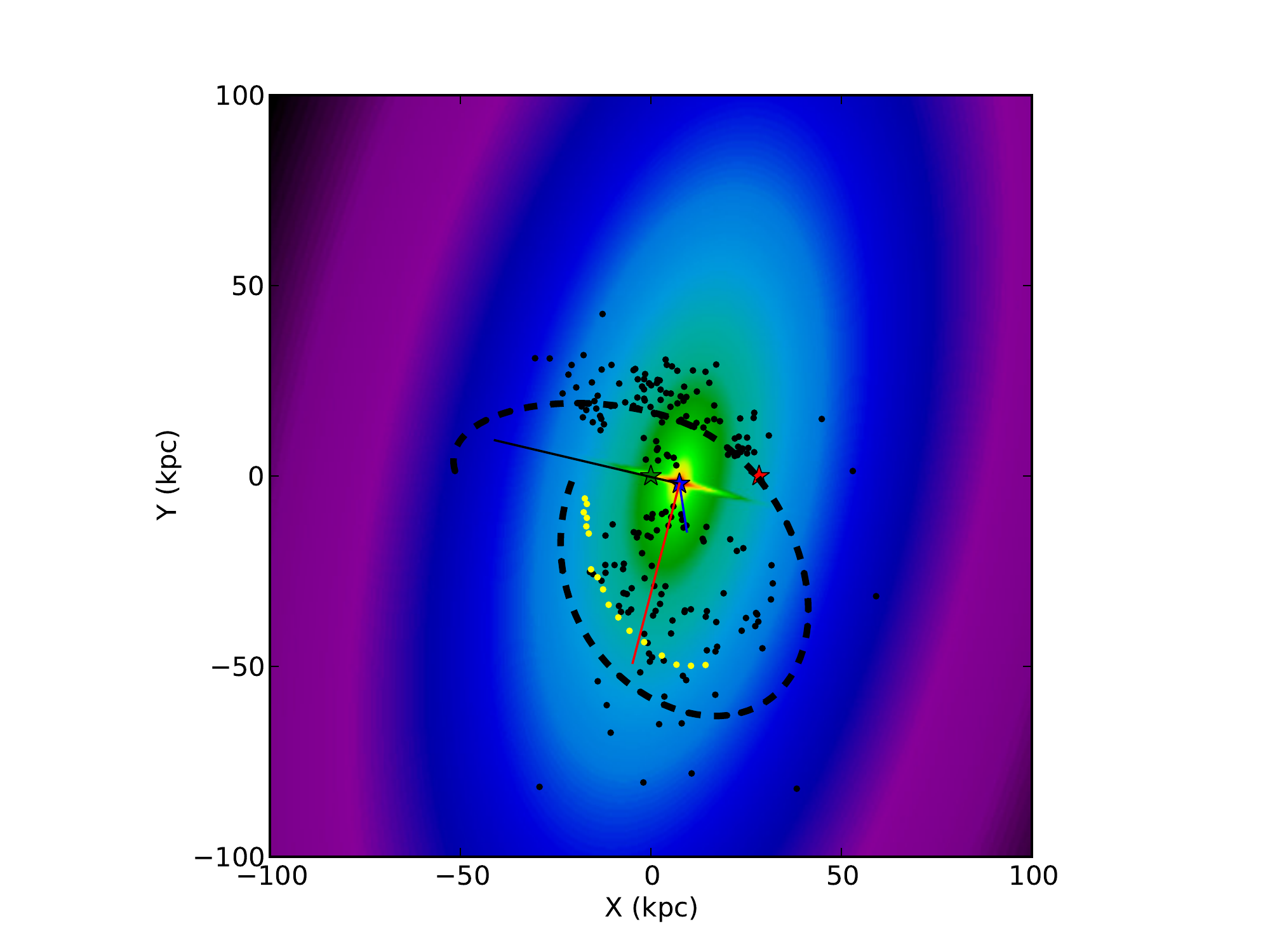} 
\caption{Total Milky Way density and the Sgr stream in the $\beta=0$
  plane.  The green, blue, and red stars show the location of the Sun,
  GC, and Sgr dwarf respectively.  The black points show the Sgr
  stream M Giants and the yellow points show the SDSS fields from
  \citet{Belokurov2006}.  The dashed curve shows the orbit of the
  dwarf in the Milky Way model with the parameters equal to the
  expectation values from Table 2.  The solid blue, red, and black
  lines show the $A$, $B$, and $1$ axes respectively.}
  \label{Fig:HaloAlign}
  \end{minipage}
\end{figure*}

\centering
\begin{figure*}
\centering
  \begin{minipage}{126mm}
 \includegraphics[width=126mm]{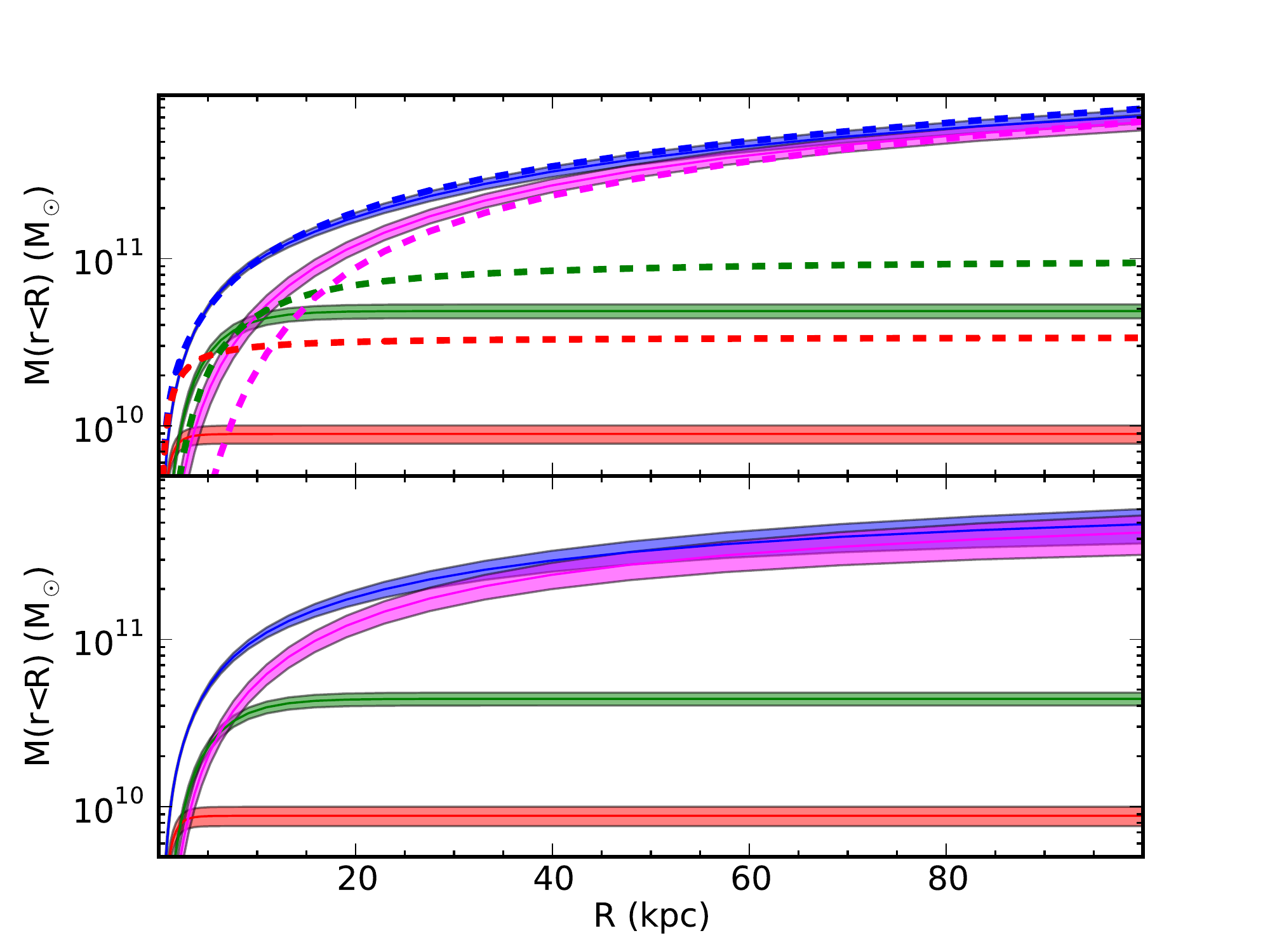} 
\caption{Cumulative mass profile as a function of spherical radius
  $r$.  Solid lines and shaded regions show the average values 68\%
  credible region for the total (blue) and the separate contribution
  of the bulge (red), disk (green), and halo (magenta).  The dashed red
  line shows the predictions of the LJM model.  The upper panel shows
  $M(r)$ with the full suite of constraints, include the data for the
  Sgr Stream.  The lower panel shows $M(r)$ when the stream data is
  not used.}
  \label{Fig:MassProfiles}
  \end{minipage}
\end{figure*}

\end{document}